\documentclass[twocolumn]{aastex7}

\shorttitle{Fundamental Stellar Parameters of Stars Central to Bowshock Nebulae}
\shortauthors{Patten et al.}
\graphicspath{{./}{figures/}}

\begin{document}

\newcommand{\moy}{$M_\odot~\text{yr}^{-1}$}
\newcommand{\um}{$\mu\text{m}$}
\newcommand{\halpha}{H-$\alpha$}
\newcommand{\mdot}{$\dot{M}$}
\newcommand{\teff}{$T_{\rm{eff}}$}
\newcommand{\logg}{$\log g$}
\newcommand{\vsini}{$v\sin i$}
\newcommand{\kms}{$\rm{km}~\rm{s}^{-1}$}
\newcommand{\degree}{$^{\circ}$}
\newcommand{\kelvin}{$K$}
\newcommand{\angstrom}{\r{A}}
\newcommand{\numstars}{$104$}
\newcommand{\numcomparison}{$28$}
\newcommand{\mstar}{$M_*$}
\newcommand{\lstar}{$L_*$}
\newcommand{\rstar}{$R_*$}
\newcommand{\numJayasinghe}{$310$}
\newcommand{\rsun}{$R_\odot$}
\newcommand{\msun}{$M_{\odot}$}
\newcommand{\lsun}{$L_{\odot}$}
\newcommand{\numBstars}{59}
\newcommand{\numOstars}{44}
\newcommand{\fracO}{43\%}
\newcommand{\fracB}{57\%}
\newcommand{\fracslow}{52\%}
\newcommand{\numslow}{54}
\newcommand{\numfast}{48}
\newcommand{\fracfast}{48~\%}
\newcommand{\numbinaries}{65}
\newcommand{\fracbinaries}{60\%}
\newcommand{\numrunaway}{19}

\title{Fundamental Parameters for Central Stars of 103 Infrared Bowshock Nebulae}

\author[0009-0005-0582-8469]{Nikhil Patten}
\affiliation{Department of Physics and Astronomy, University of Wyoming, 1000 E. University Avenue, Laramie, WY 82071, USA}
\email{npatten@uwyo.edu}

\author[0000-0002-4475-4176]{Henry A. Kobulnicky}
\affiliation{Department of Physics and Astronomy, University of Wyoming, 1000 E. University Avenue, Laramie, WY 82071, USA}
\email{chipk@uwyo.edu}

\author[0000-0001-9062-3583]{Matthew S. Povich}
\affiliation{Department of Physics and Astronomy, California State Polytechnic University, 3801 West Temple Ave, Pomona, CA 91768, USA}
\email{mspovich@cpp.edu}

\author[0000-0003-1963-5754]{Angelica S. Whisnant}
\affiliation{Department of Astronomy, Ohio State University, 4055 McPherson Laboratory, 140 West 18th Avenue, Columbus, OH, 43210, USA}
\affiliation{Department of Physics and Astronomy, California State Polytechnic University, 3801 West Temple Ave, Pomona, CA 91768, USA}
\email{whisnant.5@bucketemail.osu.edu}

\author{Sydney Andrews}
\affiliation{Department of Physics and Astronomy, Appalachian State University, 525 Rivers Street, Boone, NC 28608, USA}
\affiliation{Department of Physics and Astronomy, University of Wyoming, 1000 E. University Avenue, Laramie, WY 82071, USA}
\email{andrewssc1@appstate.edu}

\author[0009-0006-3467-630X]{Alexandra Boone}
\affiliation{Department of Physics and Astronomy, University of Wyoming, 1000 E. University Avenue, Laramie, WY 82071, USA}
\email{aboone1@uwyo.edu}

\author[0000-0003-4578-3216]{Srujan Dandu}
\affiliation{Department of Physics and Astronomy, University of Wyoming, 1000 E. University Avenue, Laramie, WY 82071, USA}
\email{sdandu@uwyo.edu}

\author{Naomi Jones}
\affiliation{Department of Physics and Astronomy, University of Wyoming, 1000 E. University Avenue, Laramie, WY 82071, USA}
\email{njones23@uwyo.edu}

\author[0009-0007-6290-3319]{S. Nick Justice}
\affiliation{Department of Physics and Astronomy, University of Wyoming, 1000 E. University Avenue, Laramie, WY 82071, USA}
\email{sjustice@uwyo.edu}

\author{Dylan Hope}
\affiliation{Department of Physics and Astronomy, California State Polytechnic University, 3801 West Temple Ave, Pomona, CA 91768, USA}
\affiliation{Department of Physics and Astronomy, University of Wyoming, 1000 E. University Avenue, Laramie, WY 82071, USA}
\email{dthope@cpp.edu}

\author[0000-0002-2401-8411]{Alexander Larsen}
\affiliation{Department of Physics and Astronomy, University of Wyoming, 1000 E. University Avenue, Laramie, WY 82071, USA}
\email{alarse15@uwyo.edu}

\author[0009-0005-1158-6777]{Ryan McCrory}
\affiliation{Department of Physics, Rhodes College, 2000 North Pkwy, Memphis, TN 38112, USA}
\affiliation{Department of Physics and Astronomy, University of Wyoming, 1000 E. University Avenue, Laramie, WY 82071, USA}
\email{mccrj-25@rhodes.edu}

\author{Julia Meredith}
\affiliation{Department of Physics, Juniata College, 1700 Moore Street, Huntingdon, PA 16652, USA}
\affiliation{Department of Physics and Astronomy, University of Wyoming, 1000 E. University Avenue, Laramie, WY 82071, USA}
\email{julia@meredithville.net}

\author{Maria Renee Meza}
\affiliation{Department of Astronomy, University of Virginia, 530 McCormick Road, Charlottesville, VA 22904, USA}
\email{qqn8hw@virginia.edu}

\author[0009-0001-2223-2975]{Alexandra C. Rosenthal}
\affiliation{Department of Astronomy and Cornell Center for Astrophysics and Planetary Science, Cornell Univsersity, Ithaca, NY 14853, USA}
\affiliation{Department of Astronomy, University of Virginia, 530 McCormick Road, Charlottesville, VA 22904, USA}
\affiliation{Department of Physics and Astronomy, University of Wyoming, 1000 E. University Avenue, Laramie, WY 82071, USA}
\email{acr6tkv@virginia.edu}

\author{William Salazar}
\affiliation{Department of Physics and Astronomy, California State Polytechnic University, 3801 West Temple Ave, Pomona, CA 91768, USA}
\affiliation{Department of Physics and Astronomy, University of Wyoming, 1000 E. University Avenue, Laramie, WY 82071, USA}
\email{wbsalazar@cpp.edu}

\author{Alexander R. Sterling}
\affiliation{Department of Physics and Astronomy, University of Notre Dame, 225 Nieuwland Science Hall, Notre Dame, IN 46556, USA}
\affiliation{Department of Physics and Astronomy, University of Wyoming, 1000 E. University Avenue, Laramie, WY 82071, USA}
\email{asterli3@nd.edu}

\author[0009-0009-4618-8049]{Noshin Yesmin}
\affiliation{Department of Astronomy, University of Virginia, 530 McCormick Road, Charlottesville, VA 22904, USA}
\email{pqt7tv@virginia.edu}

\author[0000-0002-5782-9093]{Daniel A. Dale}
\affiliation{Department of Physics and Astronomy, University of Wyoming, 1000 E. University Avenue, Laramie, WY 82071, USA}
\email{ddale@uwyo.edu}



\begin{abstract}

Stellar bowshock nebulae are arcuate shock fronts formed by the interaction of radiation-driven stellar winds and the relative motion of the ambient interstellar material. Stellar bowshock nebulae provide a promising means to measure wind-driven mass loss, independent of other established methods. In this work, we characterize the stellar sources at the center of bowshock nebulae drawn from all-sky catalogs of 24 $\mu$m-selected nebulae. We obtain new, low-resolution blue optical spectra for \numstars~stars and measure stellar parameters temperature \teff, surface gravity \logg, and projected rotational broadening \vsini. We perform additional photometric analysis to measure stellar radius \rstar, luminosity \lstar, and visual-band extinction $A_V$. All but one of our targets are O and early B stars, with temperatures ranging from $T$=16.5--46.8~k\kelvin, gravities $\log g=$2.57--4.60, and \vsini~from $<$100--400~\kms. With the exception of rapid rotator $\zeta$ Oph, bowshock stars do not rotate at or near critical velocities.  At least 60 of 103 (60\%) OB bowshock stars are binaries, consistent with the multiplicity fraction of other OB samples. The sample shows a runaway fraction of 23\%, with \numrunaway~stars having $v_{\text{2D}}\geq25$~\kms. Of the 19 runaways, at least 15 ($\geq$79\%) are binaries, favoring dynamical ejection over the binary supernova channel for producing runaways. We provide a comprehensive census of stellar parameters for bowshock stars, useful as a foundation for determining the mass-loss rates for OB-type stars---one of the single most critical factors in stellar evolution governing the production of neutron stars and black holes.

\end{abstract}



\section{Introduction} \label{sec:intro}

OB stars inject ionizing radiation and mechanical energy into their host galaxies---a process known as stellar feedback. Radiation-driven stellar winds, unique to O- and B-type stars, blow out and shape their local environments. Stellar feedback during the main-sequence and post-main-sequence stages of evolution can suppress new star formation within a few parsecs from massive stars \citep{Krumholz2007,Olivier2021}. In large groups, their combined effects on their host galaxy can even give rise to galactic-scale outflows, known as galactic fountains or chimneys \citep{Heckman1993,Dahlem1997}. At the end of the main-sequence, following significant post-main-sequence evolution, many OB stars undergo core-collapse supernovae. This process, too, can inhibit star formation as the blast waves heat and disrupt nearby molecular clouds \citep{Kortgen2016}. After their final act, OB stars leave black holes and neutron stars. This is of particular interest in the era of time-domain astronomy with the advent of gravitational wave detectors --- LIGO \citep{Abbott2009} and VIRGO \citep{Bradaschia1990} --- and proposed future missions --- LISA \citep{Arun2022} --- which allows the study of these degenerate objects and their mergers. Understanding all of these phenomena requires an understanding of the processes that govern the evolution of massive stars.

The rate at which stars lose mass by radiation-driven stellar winds is a significant factor that determines their evolution. For single-star systems, wind-driven mass loss is the primary determining factor that dictates the star's evolutionary trajectory at the end of the main-sequence. The magnitude of mass loss varies during the main- and post-main-sequence stages of evolution. Several techniques for measuring mass-loss rates have been developed, including: H$\alpha$ profile analysis \citep{KleinandCastor1978,Leitherer1988,LamersandLeitherer1993}, UV spectroscopy of metal-resonance lines \citep{Garmany1981, HowarthandPrinja1989, Fullerton2006, Marcolino2009}, and thermal radio/FIR continuum analyses \citep{Abbott1980, Bengalia2007, Massa2017}. Measured mass-loss rates $\log\dot{M}$ (\moy) can range from -4 for the most luminous O-type stars to -8 at the limit of detectability for early-B dwarfs \citep{Puls1996, Rudio-Diez2022}. Each of these techniques is subject to its own inherent limitations and uncertainties. One common source of uncertainty is the amount of clumping in the winds. Small-scale density inhomogeneities may lead to overestimates of mass-loss rates by factors of tens or hundreds if not properly taken into account \citep{Fullerton2006}. Theoretical prescriptions parameterize mass loss as a function of several key stellar parameters: luminosity $L_*$; terminal wind speed $v_\infty$; stellar mass $M_*$; effective temperature \teff; and metallicity $Z_*$ \citep{Vink2001}. Competing sets of models can reduce mass-loss predictions by factors of $2$--$3$ \citep{Bjorklund2021, Bjorklund2023, Kritcka2021}.

A new and independent method of measuring mass-loss rates uses OB stars that produce IR bowshock nebulae \citep{Kobulnicky2010, Kobulnicky2018, Kobulnicky2019}. Stellar bowshock nebulae are arc-shaped circumstellar features most prominent at mid-infrared wavelengths, sometimes also displaying emission in the optical and radio bandpasses. Stellar bowshock nebulae were first discovered by \cite{GullandSofia1979} with the detection of arc-shaped H$\alpha$ nebulae around two early-type stars. \citet{vanBuren1988} conducted the first large-scale search for extended arc- and ring-shaped features as seen at 60 and 100 \um~using Infrared Astronomical Satellite (\textit{IRAS}) survey data. These 33 newly discovered features, of which 15 resembled stellar bowshocks, were associated with hot and luminous stars. They proposed that the shock front, visible at 60 \um, is formed as a result of the collision of stellar wind from a star moving supersonically with respect to the ambient interstellar material\footnote{For completeness, other stellar sources can produce bowshock nebulae such as pulsars \citep{Hartigan1987, Helfand2001, Wang2013}, X-ray binaries \citep{Gvaramadze2011a}, low-mass evolved stars \citep{Noriega-Crespo1997b, Gvaramadze2014a, Gvaramadze2014b}, and Herbig-Haro objects \citep{Stapelfeldt1991, Chin-Fei2000, Bally2002}. These fascinating phenomena are not the subject of this paper.}. The discovery of stellar bowshock nebulae triggered a search for more of these objects in infrared wavelengths. \citet{Noriega-Crespo1997a}, examining 58 bowshock candidates selected from \textit{IRAS} survey data, were able to positively identify $\sim20$ stellar bowshocks. With the advent of Spitzer Space Telescope (\textit{SST}), \citet{Povich2008} reported the discovery of six stellar bowshock nebulae around Galactic star-forming region M17 with GLIMPSE survey data. \citet{Gvaramadze2008} and \citet{Gvaramadze2011b} discovered $3$ and $7$, respectively, bowshock nebulae by scanning 22.0 \um~Widefield Infrared Survery Explorer (\textit{WISE}) data around stars ejected from nearby associations. \citet{Kobulnicky2010} unveiled ten additional bowshock candidates in their search of \textit{SST} Cygnus X Legacy Survey data at 8 and 24 \um~bands. \citet{Peri2015} cataloged 45 bowshock nebulae revealed by \textit{WISE} data. The sample of bowshock nebulae grew to 709 in the extensive search of \textit{SST} and \textit{WISE} data in \cite{Kobulnicky2016}. This catalog of bowshock nebulae was expanded with the addition of \numJayasinghe~new candidates from a citizen-science search of \textit{SST} archival data \citep{Jayasinghe2019}. \citet{Kobulnicky2019}, following the procedure of \cite{Kobulnicky2010, Kobulnicky2018}, demonstrated the promise of using the physics and geometry of bowshock nebulae to measure mass-loss rates in their study of 70 stars at the center of bowshock nebulae, shortened to bowshock stars for the remainder of this work.


\citet{Chick2020} conducted a spectroscopic study on a subsample of $84$ bowshock stars from the \citet{Kobulnicky2016} catalog. Red-optical spectra indicated $96\%$ ($81$ of $84$) of the studied bowshock stars were early-type stars, confirming the identification of bowshock nebulae a reliable method of finding previously unknown OB stars. Their spectra allowed them to obtain rough spectral classification, with stars categorized as either O-, OB-, or B-type. However, their data precluded reliable luminosity classification. At least $\gtrsim$$36\%$ of the sampled stars also showed radial-velocity variations, indicating a significant fraction of bowshock stars were in multiple systems. \cite{KobulnickyandChick2022} studied the kinematics for $267$ bowshock stars using Gaia EDR3 proper-motion data \citep{Gaia2020}. Surprisingly, only $24\%$ were determined to be runaways, while the remaining $76\%$ of stars had 2D space velocities less than 25 \kms. This result indicated that runaway stars were not the primary driving mechanism in forming bowshock nebulae. Rather, the velocity differential generated from a bulk flow of material or interstellar density enhancements played larger-than-expected roles in the formation of stellar bowshock nebulae.


In this work, we build upon the \citet{Chick2020} study and present the results of a new spectroscopic analysis on a subsample of \numstars~bowshock stars. This work is the first comprehensive spectroscopic survey of this size for bowshock stars using blue wavelength optical-spectra that enables an improved measurement of stellar parameters \teff, \logg, and \vsini. In Section~\ref{sec:obs}, we discuss target selection from previous bowshock catalogs. In Section~\ref{sec:analysis} we present our methods of extracting stellar parameters from blue-violet optical spectra using grids of theoretical stellar spectra. In Section~\ref{sec:results} we present the results of our analysis for the \numstars~selected bowshock stars. In Section~\ref{sec:discussion} we discuss how bowshock stars compare to non-bowshock-forming OB stars. In Section~\ref{sec:conclusions}, we summarize our findings, discuss some of the implications of our work, and directions for future bowshock studies.


\section{Spectroscopic observations} \label{sec:obs}

\subsection{Target selection}


\begin{deluxetable}{ll}
	\tablecaption{Starting samples and cuts}\label{tab:cuts}
	\tablehead{K16 & 709 \\J19 & 187 of 310\\Total & 896}
	\startdata
		North of -$45^\circ$ & 589\\
		previous + $\pi/\sigma_\pi >5$ & 204 \\
		previous + $G<13.5$ & 145
	\enddata
\end{deluxetable}

We developed and applied several cuts to the \citet{Kobulnicky2016} and \citet{Jayasinghe2019} catalogs when creating our subsample of bowshock stars to observe. Starting with the 709 bowshock candidates in \citet{Kobulnicky2016} and 187 of the \numJayasinghe~reported in \citet{Jayasinghe2019} (896 in total), we selected stars north of -45\degree~declination (observable from northern-based observatories), resulting in 589 potential targets. We further selected only targets with parallax-to-parallactic-uncertainty ratios greater than 5 (i.e., reliable distances) resulting in a subsample of 204. The last cut of Gaia $G$ magnitude brighter than 13.5\footnote{This corresponds to $B_P=$14--15 mags, given typical $B_P-G$ colors of 1--2 mags.} (bright enough to perform sensitive blue-violet spectroscopy) results in a final count of 145 potential target stars. Of these, we obtained spectra for ~\numstars~candidates. Table~\ref{tab:cuts} summarizes the parent sample and the subsequent culling of the potential targets.

\begin{figure}[htb!]
        \plotone{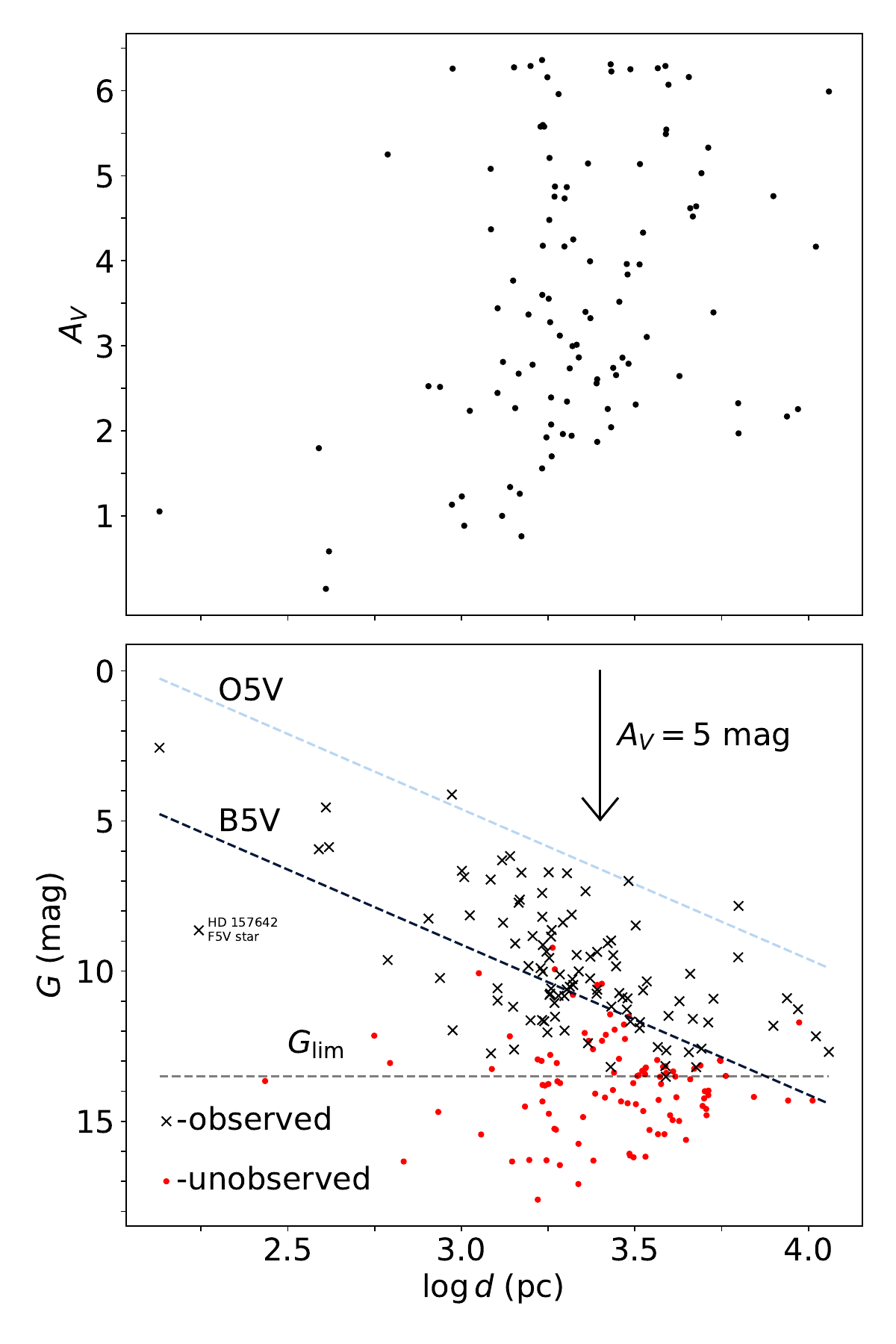}
        \caption{Completeness of bowshock sample.}
        \label{fig:completeness}
\end{figure}

Figure~\ref{fig:completeness} plots the measured visual extinction $A_V$ (described in Section~\ref{sec:SED}) against Gaia DR3 parallactic distance $d$ (\textit{top panel}) and apparent Gaia $G$-band magnitude against distance (\textit{bottom panel}) for the subsample of bowshock stars described above. In the top panel, extinction scales with distance, with scatter due to nonuniform galactic dust structure. In the bottom panel, black $\times$'s indicate the \numstars~observed program stars and red points signify unobserved bowshock candidate stars. Diagonal dashed lines indicate the unreddened apparent magnitudes of an O5 dwarf and a B5 dwarf, respectively \citep{PecautandMamajek2013}. The horizontal line illustrates the effective brightness limit of our sample of approximately $G=$13.5 magnitudes. The vertical arrow designates a typical V-band extinction of 5 magnitudes (converted to Gaia $G$-band extinction). Targets span distances ranging from 134 pc--11.5 kpc, with 11 bowshock stars within 1~kpc of the Sun.




Many targets beyond 1~kpc fall below the dim B5V track. This is expected, as distant stars lie behind large column densities of dust and can exhibit several magnitudes of extinction. It is also important to remind the reader that stars cataloged in the \citet{Kobulnicky2016} and \citet{Jayasinghe2019} works are \textit{candidate} central stars to the bowshock nebula. In some cases, the identification of the central star is ambiguous from 4.5~\um~data. For example in their spectroscopic study of bowshock stars, \citet{Chick2020} note four cases in which multiple stars reside near the geometric center of the nebula and propose alternative powering stars based on optical spectra and angular distances. For discussion of notable outlier HD 157642, see Section~\ref{app:HD157642} of the Appendix.

\startlongtable
\begin{deluxetable*}{llcccll}
	\tablecaption{Observations \label{tab:obs}}	\tablehead{
	SBN ID & Alias & R.A. (2000) & Dec. (2000) & $B_P$ & Observatory & Exposures \\
	 & &HH:MM:SS & 		$^\circ$:':'' & (mag) & & \\
	(1) &(2) &(3) &(4) &(5) &(6) &(7)}
	\startdata
		$001_\star$ & CD-29 14004 & 17:48:07.68 & -29:07:55.6 & 11.24 & APO & $3\times300$ s \\
		$007_\star$ & ... & 17:58:30.65 & -26:09:49.3 & 13.39 & APO & $1\times600$ s \\
		$010_\star$ & HD 314937 & 18:00:17.47 & -25:14:14.6 & 10.95 & APO & $2\times300$ s \\
		$013_\star$ & $\zeta$ Oph & 16:37:09.55 & -10:34:01.2 & 3.01 & APO & $1\times10$ s \\
		$016_\star$ & ... & 18:00:55.22  & -22:57:37.1 & 12.76 & APO & $1\times900$ s \\
		$019_\star$ & ... & 18:05:11.18 & -21:04:43.3 & 13.14 & APO & $1\times1200$ s \\
		$026_\star$ & ... & 18:11:59.40 & -19:36:55.4 & 12.63 & APO & $1\times150$ s \\
		$027_\star$ & TYC 6272-1457-1 & 18:12:13.03 & -19:35:24.0 & 11.24 & APO & $1\times150$ s \\
		$039_\star$ & ... & 18:08:46.42 & -16:25:59.5 & 13.01 & APO & $1\times900$ s \\
		$043_\star$ & TYC 6269-2270-1 & 18:16:25.25 & -17:05:20.8 & 11.78 & APO & $1\times900$ s \\
		$051_\star$ & NGC 6618 258 & 18:20:22.70 & -16:08:34.1 & 13.74 & APO & $2\times900$ s \\
		$054_\star$ & HD 165319 & 18:05:58.82 & -14:11:53.2 & 8.06 & WIRO & $3\times90$ s \\
		$056_\star$ & ... & 18:17:17.18 & -15:29:25.4 & 12.69 & APO & $1\times600$ s \\
		$061_\star$ & HD 168183 & 18:18:58.70 & -13:59:28.3 & 8.29 & APO & $2\times60$ s \\
		$063_\star$ & BD-13 4934 & 18:19:05.57 & -13:54:50.4 & 9.50 & APO & $2\times180$ s \\
		$064_\star$ & BD-14 5040 & 18:25:38.90 & -14:45:05.8 & 10.68 & APO & $1\times300$ s \\
		$065_\star$ & BD-13 4937 & 18:19:20.04 & -13:54:21.6 & 10.81 & APO & $1\times600$ s \\
		$066_\star$ & ... & 18:15:23.98 & -13:19:35.8 & 12.04 & APO & $1\times300$ s \\
		$067_\star$ & NGC 6611 584 & 18:18:23.64 & -13:36:28.1 & 12.33 & APO & $1\times600$ s \\
		$150_\star$ & TYC 5118-279-1 & 18:49:25.06 & -02:21:09.7 & 10.75 & APO & $1\times300$ s \\
		$200_\star$ & ... & 19:02:08.86 & 3:30:47.2 & 14.11 & APO & $2\times900$ s \\
		$240_\star$ & HD 230561 & 19:03:40.70 & 13:03:11.9 & 10.96 & APO & $1\times300$ s \\
		$255_\star$ & TYC 1054-952-1 & 19:19:00.82 & 13:42:59.0 & 11.58 & APO & $2\times360$ s \\
		$288_\star$ & HD 350123 & 19:32:44.33 & 19:58:40.8 & 11.26 & APO & $2\times360$ s \\
		$298_\star$ & HD 345117 & 19:51:08.18 & 22:49:54.5 & 9.23 & WIRO & $2\times600$ s \\
		$300_\star$ & HD 344765 & 19:44:20.06 & 23:52:45.1 & 10.91 & WIRO & $1\times600$ s \\
		$302_\star$ & LS II +23 47 & 19:45:47.52 & 24:06:00.0 & 11.12 & WIRO & $1\times600$ s \\
		$303_\star$ & ... & 19:46:06.17 & 24:11:13.9 & 14.29 & APO & $2\times1800$ s \\
		$305_\star$ & HD338936 & 19:46:22.68 & 24:37:48.0 & 10.21 & WIRO & $1\times600$ s \\
		$306_\star$ & BD+24 3883 & 19:47:21.38 & 24:33:43.9 & 10.42 & WIRO & $2\times600$ s \\
		$314_\star$ & HD 338961 & 19:50:24.41 & 27:27:55.8 & 11.04 & APO & $1\times180$ s \\
		$319_\star$ & ... & 20:05:38.11 & 36:39:38.2 & 12.87 & APO & $1\times900$ s \\
		$320_\star$ & HD 191611 & 20:09:26.06 & 36:29:19.7 & 8.66 & APO & $2\times45$ s \\
		$321_\star$ & ... & 20:14:33.29 & 36:29:49.9 & 14.06 & APO & $4\times900$ s \\
		$322_\star$ & ... & 20:14:51.10 & 36:35:58.2 & 13.51 & APO & $3\times900$ s \\
		$324_\star$ & LS II +38 19 & 20:13:28.92 & 38:14:31.6 & 11.41 & APO & $2\times45$ s \\
		$325_\star$ & HD 194303 & 20:23:35.71 & 36:55:44.0 & 8.79 & APO & $1\times240$ s \\
		$326_\star$ & TYC 2697-1046-1 & 20:32:56.26 & 36:12:00.4 & 11.37 & APO & $1\times300$ s \\
		$328_\star$ & ... & 20:35:34.44 & 36:34:25.3 & 12.55 & APO & $2\times45$ s \\
		$330_\star$ & HD 229159 & 20:22:54.02 & 39:12:28.8 & 8.78 & APO & $2\times240$ s \\
		$331_\star$ & LS II +39 53 & 20:27:17.57 & 39:44:32.6 & 10.55 & APO & $2\times300$ s \\
		$332_\star$ & ... & 20:26:24.89 & 40:01:41.2 & 12.76 & APO & $3\times480$ s \\
		$335_\star$ & TYC 3156-1106-1 & 20:28:15.38 & 40:44:04.6 & 11.51 & APO & $2\times360$ s \\
		$339_\star$ & ... & 20:36:04.51 & 40:56:13.2 & 13.01 & APO & $2\times500$ s \\
		$340_\star$ & HD 195229 & 20:28:30.24 & 42:00:35.3 & 7.68 & APO & $2\times45$ s \\
		$341_\star$ & ... & 20:34:28.94 & 41:56:16.8 & 13.36 & APO & $5\times900$ s \\
		$343_\star$ & ... & 20:30:34.97 & 44:18:54.7 & 7.25 & APO & $1\times45$ s \\
		$344_\star$ & BD+43 3654 & 20:33:36.07 & 43:59:07.4 & 7.58 & APO & $2\times300$ s \\
		$345_\star$ & HD 199021 & 20:52:53.21 & 42:36:27.7 & 8.58 & APO & $1\times180$ s \\
		$346_\star$ & ... & 20:51:50.21 & 43:19:29.6 & 13.70 & APO & $3\times900$ s \\
		$349_\star$ & LS III +49 20 & 21:21:39.07 & 50:02:01.0 & 11.37 & APO & $1\times900$ s \\
		$350_\star$ & ... & 21:22:55.70 & 50:24:25.2 & 13.35 & APO & $3\times480$ s \\
		$351_\star$ & TYC 3979-1998-1 & 22:12:30.07 & 55:32:07.4 & 11.49 & APO & $2\times600$ s \\
		$353_\star$ & ... & 22:16:32.93 & 59:23:30.8 & 12.36 & APO & $2\times900$ s \\
		$354_\star$ & ... & 22:27:41.42 & 57:41:04.6 & 12.49 & APO & $2\times900$ s \\
		$356_\star$ & HD 240015 & 22:39:17.71 & 59:01:00.5 & 10.18 & APO & $2\times300$ s \\
		$357_\star$ & HD 240016 & 22:39:22.54 & 59:00:25.9 & 9.84 & APO & $1\times300$ s \\
		$358_\star$ & TYC 4264-1036-1 & 22:34:06.50 & 60:58:50.2 & 10.65 & APO & $2\times300$ s \\
		$359_\star$ & HD 215806 & 22:46:40.22 & 58:17:43.8 & 9.30 & APO & $1\times300$ s \\
		$360_\star$ & HD 216411 & 22:51:33.77 & 59:00:30.6 & 7.35 & APO & $2\times30$ s \\
		$361_\star$ & TYC 4278-522-1 & 22:51:39.72 & 61:08:51.0 & 10.89 & APO & $2\times600$ s \\
		$362_\star$ & ... & 22:55:45.4 & 60:24:40.3 & 13.69 & APO & $6\times900$ s \\
		$364_\star$ & BD+60 39 & 00:21:53.88 & 61:45:02.5 & 9.49 & APO & $1\times600$ s \\
		$365_\star$ & HD 2619 & 00:30:28.32 & 65:16:19.6 & 8.43 & APO & $3\times100$ s \\
		$366_\star$ & HD 2083 & 00:25:51.24 & 71:48:25.6 & 6.85 & APO & $2\times30$ s \\
		$367_\star$ & TYC 4046-1614-1 & 02:17:53.21 & 61:11:12.8 & 11.16 & APO & $2\times360$ s \\
		$368_\star$ & V* KM Cas & 02:29:30.46 & 61:29:44.2 & 11.49 & APO & $2\times360$ s \\
		$369_\star$ & BD+60 586 & 02:54:10.68 & 60:39:03.6 & 8.52 & APO & $1\times180$ s \\
		$371_\star$ & TYC 3339-851-1 & 04:05:53.04 & 51:06:58.0 & 11.91 & APO & $1\times600$ s \\
		$372_\star$ & TYC 3340-2437-1 & 04:11:10.73 & 50:42:29.5 & 11.55 & APO & $1\times600$ s \\
		$373_\star$ & HD 21856 & 03:32:40.01 & 35:27:42.1 & 5.82 & APO & $2\times60$ s \\
		$374_\star$ & HD 41161 & 06:05:52.44 & 48:14:57.5 & 6.66 & APO & $3\times45$ s \\
		$375_\star$ & V* AE Aur & 05:16:18.14 & 34:18:45.0 & 6.08 & APO & $1\times30$ s \\
		$376_\star$ & HD 48099 & 06:41:59.23 & 06:20:43.4 & 6.27 & APO & $1\times30$ s \\
		$377_\star$ & HD 46573 & 06:34:23.57 & 02:32:03.1 & 8.00 & APO & $1\times120$ s \\
		$378_\star$ & $\upsilon$ Ori & 05:31:55.85 & -07:18:05.8 & 4.55 & APO & $1\times15$ s \\
		$381_\star$ & HD 54662 & 07:09:20.26 & -10:20:47.8 & 6.16 & APO & $1\times45$ s \\
		$383_\star$ & CD-26 5136 & 07:53:01.01 & -27:06:57.6 & 9.75 & APO & $1\times360$ s \\
		$384_\star$ & CD-35 4415 & 08:16:32.09 & -35:38:52.8 & 10.72 & APO & $1\times550$ s \\
		$385_\star$ & CD-41 4637 & 08:55:27.67 & -41:35:22.2 & 9.95 & APO & $1\times720$ s \\
		$386_\star$ & V* GP Vel & 09:02:06.86 & -40:33:16.9 & 7.00 & APO & $3\times45$ s \\
		$388_\star$ & HD 75860 & 08:50:53.23 & -43:45:05.4 & 7.79 & APO & $1\times180$ s \\
		$389_\star$ & HD 76031 & 08:52:04.13 & -44:00:34.9 & 9.11 & APO & $1\times600$ s \\
		$634_\star$ & HD 152756 & 16:57:15.00 & -43:43:15.2 & 9.19 & APO & $1\times300$ s \\
		$640_\star$ & HD 326533 & 16:58:24.05 & -42:43:41.2 & 10.23 & APO & $1\times300$ s \\
		$662_\star$ & HD 153426 & 17:01:13.01 & -38:12:11.9 & 7.46 & APO & $1\times120$ s \\
		$667_\star$ & V* V1012 Sco & 17:15:22.32 & -38:12:46.8 & 6.64 & APO & $1\times120$ s \\
		$673_\star$ & ... & 17:20:38.23 & -38:01:48.7 & 14.71 & APO & $2\times1200$ s \\
		$687_\star$ & ... & 17:28:59.78 & -36:06:54.7 & 14.41 & APO & $2\times1200$ s \\
		$692_\star$ & ... & 17:27:11.23 & -34:14:35.2 & 11.73 & APO & $3\times300$ s \\
		$700_\star$ & ... & 17:33:47.88 & -31:16:27.1 & 14.30 & APO & $2\times1200$ s \\
		$705_\star$ & CD-29 13925 & 17:45:08.21 & -29:56:45.2 & 10.93 & APO & $1\times600$ s \\
		2G0063185-0066622$_\star$ & ... & 18:02:29.23 & -23:49:34.0 & 12.78 & APO & $2\times500$ s \\
		2G0065094-0061344$_\star$ & ... & 18:02:41.88 & -23:38:02.0 & 12.84 & APO & $2\times900$ s \\
		2G0118310-0062990$_\star$ & HD 166965 & 18:13:51.38 & -18:59:22.2 & 8.78 & APO & $2\times120$ s \\
		2G0592097+0012519$_\star$ & ... & 19:41:43.08 & 23:16:01.9 & 12.87 & APO & $2\times400$ s \\
		2G0600615-0018340$_\star$ & HD 344878 & 19:44:44.06 & 23:51:07.9 & 10.72 & APO & $2\times300$ s \\
		2G0770694+0193568$_\star$ & HD 193427 & 20:18:45.17 & 39:24:22.3 & 9.31 & APO & $2\times300$ s \\
		2G1045666+0128085$_\star$ & BD+57 2513 & 22:22:02.57 & 58:43:09.5 & 9.76 & APO & $1\times300$ s \\
		... & $\kappa$ Cas & 00:33:00.00 & 62:55:54.5 & 4.17 & APO & $2\times10$ s \\
		2G3525610+0001887$_\star$ & CD-35 11561 & 17:26:36.10 & -35:11:32.3 & 11.97 & APO & $1\times600$ s \\
		2G3527642+0032660$_\star$ & HD 157642 & 17:25:54.60 & -34:51:05.8 & 8.91 & APO & $1\times120$ s \\
		2G3533026+0008095$_\star$ & HD 319881 & 17:28:21.67 & -34:32:30.5 & 10.32 & APO & $1\times600$ s \\
		2G3585210+0089852$_\star$ & ... & 17:38:32.28  & -29:43:07.0 & 12.69 & APO & $1\times60$ s \\
	\enddata
	\tablecomments{Columns: (1) Stellar bowshock nebula identifier from \cite{Kobulnicky2016} and addendum \citet{Jayasinghe2019} catalogs; (2) Common alias, if available; (3) Right Ascension, in HH:MM:SS; (4) Declination, in DD:MM:SS; (5) Gaia $B_P$ magnitude; (6) Observatory used, either APO or WIRO (7) Number of exposures and the length of the exposures.}
\end{deluxetable*}

Table~\ref{tab:obs} presents an overview of the \numstars~bowshock stars observed. Column~1 is the identification number of the bowshock nebula from the 709 objects cataloged in \citet{Kobulnicky2016} and additional \numJayasinghe~nebulae from \citet{Jayasinghe2019}. Column~$2$ is a common alias of the star, where available. Columns~$4$ and $5$ are the right ascension and declination of the program stars in J2000 coordinates. Column~$5$ is the Gaia $B_P$ magnitude of the star. Column~$6$ indicates the observatory used. Most of the stars were observed at APO, but six stars were observed from WIRO. Column~$7$ provides the number and duration of exposures.

\begin{figure}[htb!]
        \plotone{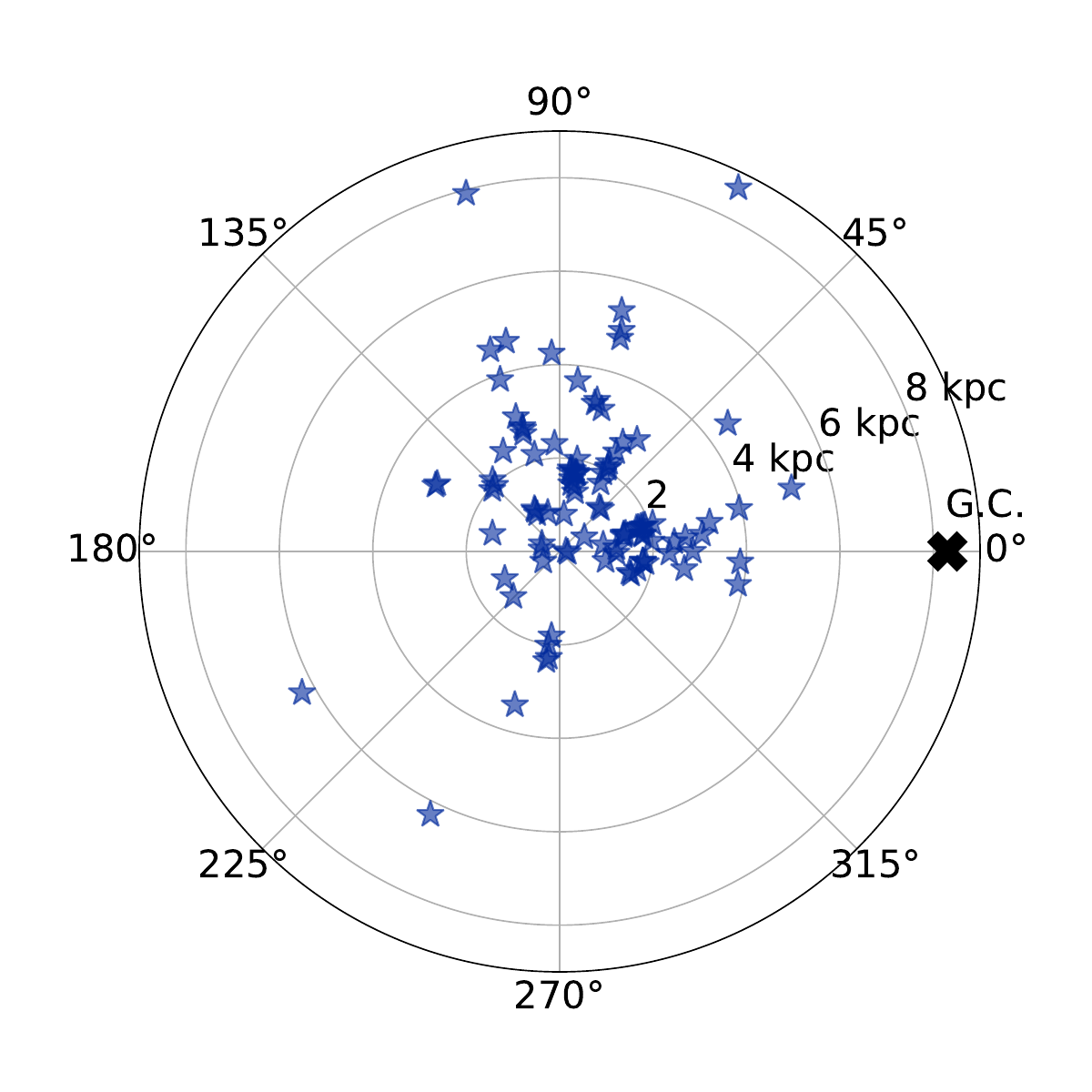}
        \epsscale{1}
        \caption{Observed program stars (indicated by blue stars) plotted on the sky.}
        \label{fig:targs}
\end{figure}

The target stars reside almost entirely within $\pm1^\circ$ of the galactic plane \citep[see Figure 3 of][]{Kobulnicky2016}, with notable exception of the nearby bowshock star $\zeta$~Oph lying 24\degree~above the plane. Figure~\ref{fig:targs} plots the Galactic longitude of the \numstars~target stars in a polar representation with the Sun at the center and Galactic center towards the right. The radial distance from the center indicates heliocentric distance, derived from Gaia EDR3 inverse parallaxes. Distances range from 134~pc -- 11.5~kpc, with a median distance of 2~kpc. For viewing purposes, the plot in Figure~\ref{fig:targs} is limited to only show targets within 9~kpc.

We performed observations of \numcomparison~well-studied, non-program OB stars. These stars, referred to hereafter as the comparison sample, were observed to compare the retrieved stellar parameters in this work to those in other published works. We selected them on the basis of being OB-type stars with previously measured values for stellar parameters (\teff, \logg, and \vsini) as well having bright apparent magnitudes ($B_P<10$~mags). As a result of these selection criteria, the comparison sample largely overlaps with stars cataloged in the \texttt{IACOB} project \citep{Simon-Diaz2011}.

\subsection{Spectrograph configuration}\label{sec:specconfig}



Observations were carried out at Apache Point Observatory (APO) with the 3.5 m telescope and at Wyoming Infrared Observatory (WIRO) with the 2.3 m telescope. Observations at APO were performed with the KOSMOS spectrograph. At APO, we used the blue 704 lines mm$^{-1}$ second order grism, yielding a reciprocal dispersion of $0.68$~\angstrom~pixel$^{-1}$. The $0''.83\times360''$ slit produced spectra with resolution $R=2,200$ and spectral coverage $\lambda\lambda3700$--$6200$~\angstrom\footnote{On a few nights, the $0''.87$ slit was used, yielding similar resolution and wavelength coverage.}. The longslit spectrograph at WIRO used a 600 lines mm$^{-1}$ grating in second order with the $1''.2\times120''$ slit. This configuration yielded a reciprocal dispersion of 1.10 \angstrom~pixel$^{-1}$, resolution $R=$1,500, and coverage of $\lambda\lambda4000$--$5800$~\angstrom.

We performed wavelength calibrations with the internal Kr and Ar calibration lamps at APO, yielding solutions with an RMS of 0.06~\angstrom, and with the intenral CuAr calibration lamp at WIRO, providing solutions with an RMS of 0.05~\angstrom. We obtained multiple comparison-lamp exposures throughout the night, and applied an interpolated wavelength solution to science frames to remove the effect of instrument flexure on the wavelength solutions. We divided science images in \texttt{iraf} by flat fields obtained from quartz dome lamp exposures. Spectra were then normalized and combined using tasks in the \texttt{specred} package. At APO, exposure times of 2$\times$45s and 2$\times$1200s for $B_P=$8.7--14.3 mag targets yielded spectra with continuum signal-to-noise ratios (SNR) of 90:1 and 30:1 pixel$^{-1}$ at $4200$~\angstrom, respectively. SNR was heavily dependent on seeing conditions and 2--3 spectra were often obtained in the same night and combined in the reduction process to maximize the SNR. We calculated Heliocentric offsets and shifted spectra to the that frame of reference. Examining a subsample of 13 stars from the comparison sample, we compared our derived radial velocities to those found in other works. Our radial velocities are scattered about the literature values with an RMS of 30~\kms. Owing to the low $R$ of the instrument, we do not attempt to measure radial velocities to a high degree of precision.



\begin{figure*}[htb!]
        \plotone{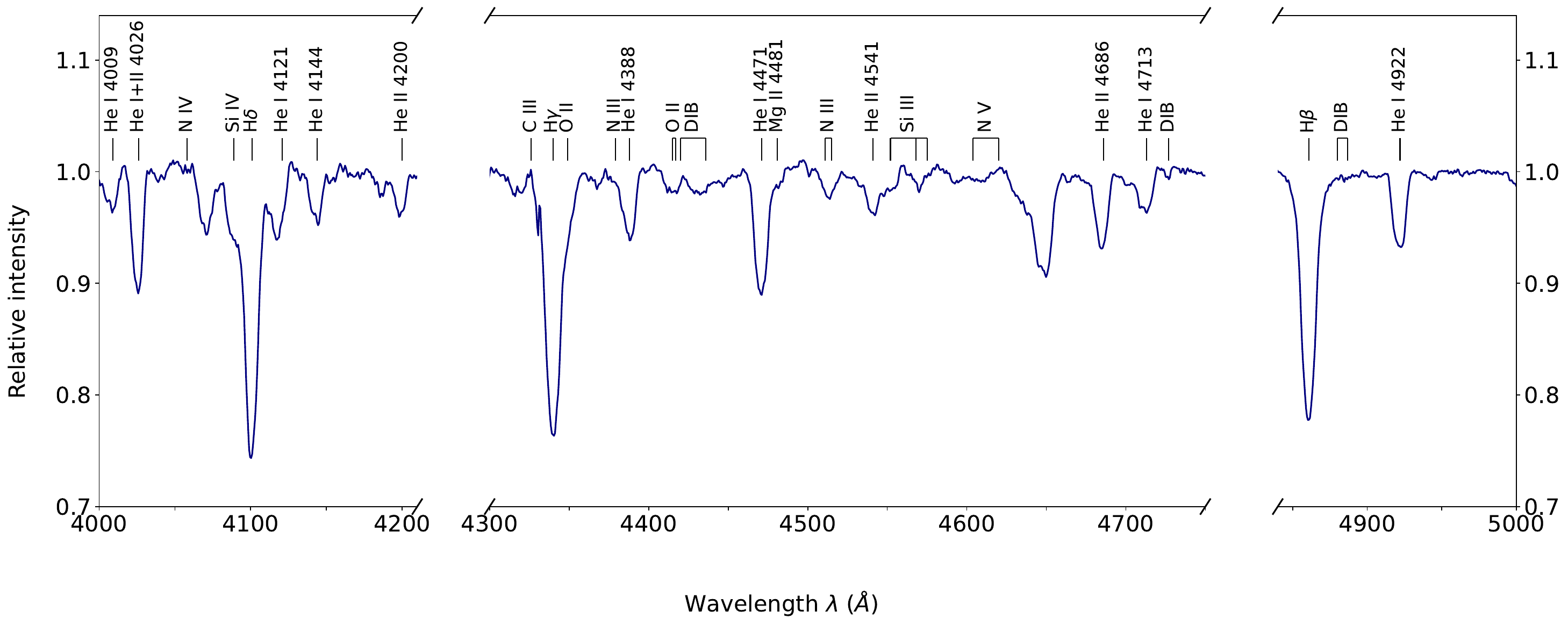}
        \caption{APO spectrum of $\zeta$ Oph.}
        \label{fig:examp_spec}
\end{figure*}

Figure~\ref{fig:examp_spec} presents the spectrum of bright bowshock nebula central star $\zeta$ Ophiuchi --- identifier BS013 in the \citet{Kobulnicky2016} catalog --- with normalized intensity plotted against wavelength. Atomic transitions sensitive to changes in stellar parameters (\teff~and \logg) in the regime of OB stars are labeled. All program stars exhibit strong Balmer features and \ion{He}{1} transitions, as is typical for OB stars. Many bowshock stars also express absorption in the \ion{He}{2} Pickering series, indicating O-type stars. Weaker blended groups of ionized metals, sensitive to changes in luminosity and temperature, are also apparent. Ratios of \ion{He}{2}/\ion{He}{1}, \ion{Si}{4}/\ion{Si}{3}, and \ion{Si}{3}/\ion{Si}{2} are good indicators of temperature \teff, and ratios of \ion{Si}{4}/\ion{He}{1}, \ion{Si}{3}/\ion{He}{1} gauge luminosity \lstar~\citep{Gray2009}. Diffuse interstellar absorption bands (DIBs) are present at $\lambda4430$~\angstrom~and $\lambda 4727$~\angstrom, to varying degrees, in all target spectra. Section~\ref{sec:appendix-spectra} in the Appendix contains vertically stacked normalized spectra for all \numstars~target stars, in order of \citet{Kobulnicky2016} identifier number for stars drawn from the \citet{Kobulnicky2016} catalog, and in order of galactic longitude for stars chosen from the \citet{Jayasinghe2019} catalog.

\section{Analysis}\label{sec:analysis}

\subsection{Temperature, gravity, and rotation from spectral fitting} \label{sec:spectralfitting}

We extracted stellar parameters from spectra with a custom python grid-search fitting routine. This code compares stellar spectra to a grid of model synthetic spectra from the TLUSTY OSTAR2002 \citep{LanzandHubeny2003} and BSTAR2006 \citep{LanzandHubeny2007} libraries, smoothed and interpolated to match the resolution of the data. TLUSTY model spectra spanned \teff~from 15--30~k\kelvin~in steps of 1~k\kelvin~for the BSTAR2006 models and 27.5--55~k\kelvin~in steps of 2.5~k\kelvin~for the OSTAR2002 models. Surface gravities \logg~ranged ~from 1.75--4.75 in steps of 0.10 for the BSTAR2006 set and 3.0--4.75 in steps of 0.25 for the OSTAR 2002 models. We convolved model spectra with instrumental and additional rotational broadenings ranging from 10--500~\kms~using the \texttt{rotin} IDL code \citep{HubenyandLanz2011}. Our spectral-fitting code determines a best-fit model spectrum by comparing the residuals between the input spectrum and all synthetic spectra, and selecting the model with the smallest absolute deviation as the best fit. The stellar parameters from the best-fit model spectrum were then used as priors in the next step of the analysis.

After this rudimentary fitting routine, we performed a Bayesian MCMC analysis. MCMC fitting for spectral analysis requires a full characterization of uncertainties. The most eminent contributor to uncertainty in the observed spectrum is the statistical noise of the spectrum, indicated by the SNR at any given pixel. We measured this quantity by following the procedure described in \citet{Irrgang2014}. We created a histogram of the dispersion, $\Delta_i$, which we define as

\begin{equation}
    \Delta_i \equiv f_i -\frac{1}{2}f_{i-2} -\frac{1}{2}f_{i+2}\text{,}
\end{equation}

\noindent for all $i$ pixels, with $f_i$ representing the continuum-normalized intensity at the $i$th pixel. The quantity $\Delta_i$ represents the equally weighted deviation between nearby pixels separated by two pixels, since adjacent pixels are likely to be correlated. The standard deviation of the $\Delta_i$ histogram was used as the empirical statistical uncertainty for the spectrum. Another important source of uncertainty is the systematic uncertainty represented by features in the stellar spectrum that are not present in the model. The source of these systematic differences can be interstellar features (DIBs and \ion{Na}{1} doublet at $\lambda5890$ and $\lambda5896$ for example), element enhancements from mass transfer and mergers for binary systems \citep{Langer2012,deMink2013}, mixing processes \citep{Rivero-Gonzalez2012,Carneiro2016,Grin2017}, degree of broadening due to micro- and macro-turbulence \citep{Villamariz2000, Ryans2002, Markova2008}, or stellar wind phenomena not included in the TLUSTY models. To obtain an estimation of these systematics, we used the residuals from the best-fit model spectrum and smoothed this residual spectrum by three pixels. At each pixel, we computed a complete uncertainty estimation by adding the statistical uncertainty component in quadrature with the systematic uncertainty, as described in \citet{Czekala2015}. With an empirical estimation of uncertainty at each pixel, the MCMC walkers explored the 3D parameter space of stellar parameters (i.e., \teff, \logg, and \vsini). At each walker step, our code computes a model spectrum using linear interpolation from the model grid, and denotes an associated reduced $\chi^2_{\text{red}}$ for this interpolated model, given by the equation:

\begin{equation}
    \chi^2_{\text{red}} = \frac{1}{n_{\text{pix}}}\sum_{i=0}^{n_{\text{pix}}}\frac{\left(f_{\text{data}, i}-f_{\text{model}, i}\right)^2}{\sigma_i}\text{ ,}
\end{equation}

\noindent where $f_{\text{data}, i}$ signifies the normalized intensity, $f_{\text{model}, i}$ the normalized intensity of the model spectrum, and $\sigma_i$ the uncertaintiy at that pixel, for all $i$ pixels between $\lambda\lambda$4000--5000.

\begin{figure}[htb!]
        \plotone{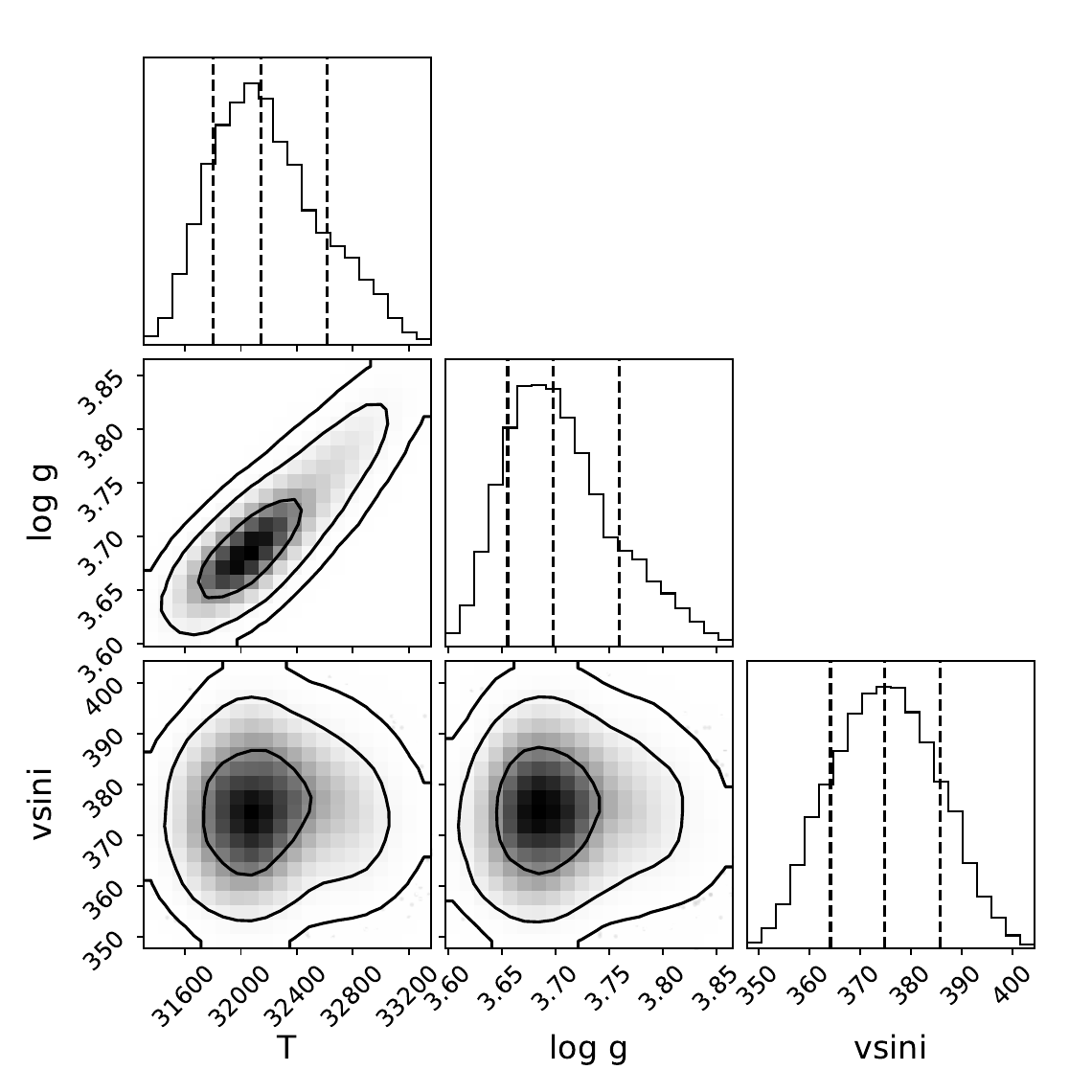}
        \caption{Posterior distribution on stellar parameters.}
        \label{fig:posteriors}
\end{figure}

Figure~\ref{fig:posteriors} plots the posterior distribution of the stellar parameters for the bright O8-9 V star $\zeta$ Oph. The shape of the posteriors indicates that the fitted parameters are well-constrained with a well-defined preferred solution. The median values for temperature, \logg, and \vsini~ of 32.1~k\kelvin, 3.70, and 375~\kms~compare favorably with the reported values of 32.0~k\kelvin, 3.71, and 385~\kms~\citep{Holgado2022}. The widths of the posterior distributions are measures of uncertainty on each parameter. We chose the 16th- and 84th-percentile values in the explored samples as the 1$\sigma$ uncertainties. The uncertainties in the fitted stellar parameters (500~\kelvin\ in \teff, 0.06 in \logg, and 11~\kms\ for the test case $\zeta$ Oph) represent the range of stellar parameters explored by the MCMC walkers in determining the best fit stellar model.

\begin{figure*}[htb!]
        \plotone{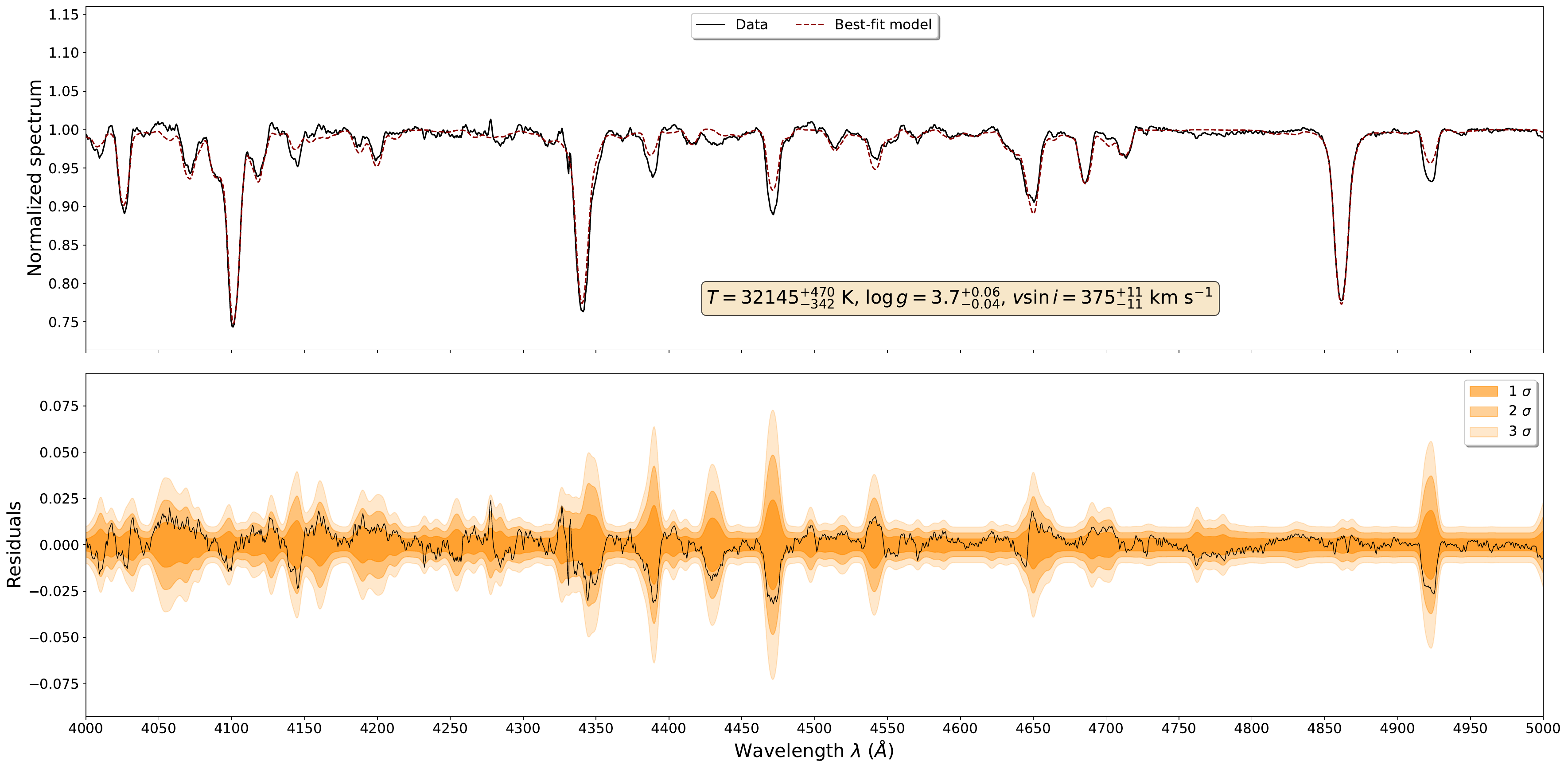}
        \caption{Top: Observed spectrum (black) overplotted with best-fit, interpolated model (dashed-red). Bottom: Residual spectrum (black) with the uncertainties at each wavelength (orange).}
        \label{fig:fitted_spectrum}
\end{figure*}

Figure~\ref{fig:fitted_spectrum} shows the results of our fitting process on the spectrum of $\zeta$ Oph. The top panel demonstrates the agreement between the input spectrum (\textit{black line}) and the best-fit interpolated model (\textit{dashed red line}). The bottom panel plots the residuals (\textit{black line}) at each pixel and the 1, 2, and 3$\sigma$ uncertainty levels (in decreasing opacities---\textit{orange}) at each pixel. The 1, 2, and 3$\sigma$ uncertainty spectrum displayed in the bottom panel illustrates the uncertainty at each pixel the MCMC code used when evaluating the $\chi^2_{\text{red}}$ for each sampled model. The residual spectrum is almost completely enveloped by the 1$\sigma$ uncertainty spectrum, indicating that the uncertainty spectrum is appropriately characterized. The largest deviations between the data and model are within a few percent. Differences in the continuum regions are random and encompassed fully by the statistical noise. Systematic features in the residuals lie at the wavelengths of stellar atmospheric lines that are not well-fit by any model. For $\zeta$ Oph in particular, these systematic features are predominantly He lines, with the EW observed in the spectrum greater than in the models. An explanation for this mismatch in the case of $\zeta$ Oph could be the star's formation history. \citet{vanRensbergen1996} propose that $\zeta$ Oph is a former secondary member in a previous binary system. Examining proper motion and evolutionary data, \citet{vanRensbergen1996} hypothesized that mass transfer from the primary deposited enriched material onto $\zeta$ Oph and spun up the star, creating the He-enriched rapid rotator we observe today. The TLUSTY models are computed with solar $Y_{\text{He}}=$0.10, as opposed to the measured $Y_{\text{He}}\approx$0.20 for $\zeta$ Oph as found in other works \citep{HowarthandSmith2001,Shepard2022, deBurgos2024}. In spite of this, solar abundances are a reasonable assumption given the unknown histories of each of the bowshock stars. Furthermore,  metal- and helium-abundances are not of interest in this study, nor are they readily measurable from low-resolution spectra.


\subsection{Mass and radius from SED fitting}\label{sec:SED}

Stellar spectra provide information on stellar parameters temperatire \teff, surface gravity \logg, and projected rotational broadening \vsini. Since \logg~is dependent on both stellar mass \mstar~and stellar radius \rstar, analysis of the stellar spectra cannot explicitly constrain these parameters. These additional parameters can be obtained by fitting stellar models to broadband, multiwavelength, photometric data, given Gaia EDR3 parallaxes and stellar spectral parameters \teff~and \logg~as priors. We retrieved photometric data from several astronomical surveys in different wavelength regimes. In the optical, we used Gaia EDR3 $G_P$, $R_P$, and $B_P$ and APASS \citep{Henden2016} U, B, V, u', g', and r' magnitudes. In the near infrared, we utilized J, H, and K magnitudes from 2MASS and W1 and W2 magnitudes from \textit{WISE}. We used EXOFASTv2 \citep{Eastman2019} with the MIST \citep{Choi2016} evolutionary models to obtain fits for stellar radius, evolutionary mass, and visual extinction $A_V$.


\subsection{MK spectral typing}

In addition to measuring fundamental stellar parameters temperature \teff, surface gravity \logg, and projected rotational velocity \vsini, from blue-violet optical spectra, as well as stellar mass \mstar, stellar radius \rstar, luminosity \lstar, and visual-band extinction $A_V$ from SED fitting, we also report approximate MK spectral types for all \numstars~bowshock stars based on visual inspection of spectra. We used the categorization scheme tabulated in the work of \citet{Liu2019}, and spectral peculiarities explained in \citet{Sota2011}. Our spectral types are approximate and are not representative of the complexity of the full MK spectral classification scheme. We report them merely for comparison purposes. For more rigorous works on the MK classification of O and B stars, see \citet{Sota2011} and \citet{Negueruela2024}.

\subsection{Binarity indicators}

Due to the large multiplicity fraction of early-type stars \citep{Sana2011, Chini2012, Kobulnicky2014}, we expect some fraction of bowshock stars to exhibit binarity. In this work we designate single line spectroscopic binaries with the flag ``SB1'', as indicated by radial velocity variations exceeding those expected from a single star  \citep[$P_{\text{single}}\leq0.05$;][]{KobulnickyandChick2022}\footnote{This flag is only applicable for the subset of stars that are studied in both this work and \citet{KobulnickyandChick2022}.}. We use ``SB2'' to flag double-lined spectroscopic binaries, as indicated either by line splitting in our spectra or spectra with unusual line ratios indicating composite spectra produced by two stars of different temperatures (i.e., strong \ion{He}{2} lines in conjunction with \ion{Mg}{2} features). We use ``EB'' to flag bowshock stars that are found to be eclipsing binaries in \citet{Malkov2006}. We flag probable astrometric binaries with ``R'', denoting a Gaia RUWE value $>1.4$\footnote{Owing to the imprecision of astrometric solutions for bright stars, we exclude stars brighter than Gaia $B_P<8$ mags from this criterion.}. In our analysis, we count stars with one or more of these flags as binaries. For probable binaries, we caution against taking the derived stellar parameters at face value, as our analysis is predicated on the assumption of one star. A detailed spectral decomposition analysis of composite spectra is infeasible for most of our targets given the low resolution and limited SNR.  

\subsection{Comparison sample results}\label{sec:compresults}

\subsubsection{Spectral fitting comparison}\label{sec:speccomp}

\startlongtable
\begin{deluxetable*}{lllccccccl}
	\tablecaption{Comparison Results\label{tab:comp}}
	\tablehead{
	ID & R.A. (2000) & Dec. (2000) & $T_{\text{eff}}$ & $\sigma_{T_{\text{eff}}}$ & $\log g$ & $\sigma_{\log g}$ & \vsini & $\sigma_{v\sin i}$  & Reference(s)\\
	 &(HH:MM:SS) & 		($^\circ$:':'') & (\kelvin) &(\kelvin) & & & (\kms) & (\kms) \\
	(1) &(2) &(3) &(4) &(5) &(6) &(7) &(8) &(9) &(10)}
	\startdata
		HD 36371 &05:32:43.67 &32:11:31.28 &14600& 300& 2.11& 0.06 & 36& 5& W22\\
		&&& 16100 & 600 & 2.21 &0.08 & 33 & 22 & TW\\
		HD 51309 &06:56:08.22 &-17:03:15.26 &15600& 400& 2.59& 0.05 & 30& 6& W22\\
		&&& 17100 & 500 & 2.64 &0.07 & 26 & 18 & TW\\
		HD 45418 &06:27:00.88 &-04:21:20.33 &16800& 130& 4.27& 0.04 & 247& 8& M15\\
		&&& 15500 & 300 & 3.86 &0.05 & 197 & 21 & TW\\
		HD 29309 &04:38:15.23 &31:59:55.64 &16800& 300& 3.70& 0.04 & 38& 12& H10\\
		&&& 16400 & 600 & 3.55 &0.08 & 41 & 30 & TW\\
		HD 47240 &06:37:52.70 &04:57:24.00 &19500& ...& 2.47& ...& 111& ...& S17\\
		&&& 22600 & 1100 & 2.93 &0.11 & 101 & 23 & TW\\
		HD 35708 &05:27:38.08 &21:56:13.09 &20700& 200& 4.15& 0.07 & 25& 2& N12\\
		&&& 19300 & 700 & 3.85 &0.08 & 30 & 21 & TW\\
		HD 30677 &04:50:03.61 &08:24:28.24 &21900& 219& 2.56& 0.03 & 168& 1& B23\\
		&&& 22200 & 800 & 2.92 &0.07 & 147 & 15 & TW\\
		HD 202347 &21:13:47.86 &45:36:41.26 &22600& 125& 3.77& 0.04 & 95& 5& M15\\
		&&& 21900 & 700 & 3.83 &0.09 & 117 & 21 & TW\\
		HD 34989 &05:21:43.56 &08:25:42.80 &24800& 450& 4.19& 0.04 & 30& 16& H10\\
		&&& 25800 & 1200 & 3.97 &0.10 & 27 & 19 & TW\\
		BD+35 3956 &20:05:59.98 &35:45:44.42 &24800& ...& ...& ...& 330& 10& D07; H10\\
		&&& 23300 & 700 & 3.52 &0.08 & 320 & 16 & TW\\
		HD 192660 &20:14:26.08 &40:19:45.05 &25200& 130& 2.90& 0.04 & 145& 5& M15\\
		&&& 26300 & 1000 & 3.03 &0.11 & 117 & 24 & TW\\
		HD 225146 &00:03:57.50 &61:06:13.09 &28300& ...& 3.11& ...& 67& ...& H22\\
		&&& 29000 & 800 & 3.37 &0.12 & 102 & 20 & TW\\
		HD 47432 &06:38:38.19 &01:36:48.68 &29100& ...& 3.04& ...& 97& ...& H22\\
		&&& 28500 & 500 & 3.08 &0.05 & 115 & 24 & TW\\
		HD 37128 &05:36:12.81 &-01:12:06.91 &29500& ...& 3.22& ...& 55& ...& S17\\
		&&& 26700 & 900 & 2.97 &0.08 & 32 & 21 & TW\\
		HD 61347 &07:38:16.12 &-13:51:01.22 &30900& ...& 3.12& ...& 112& ...& S17\\
		&&& 32500 & 600 & 3.46 &0.06 & 95 & 38 & TW\\
		HD 149757 &16:37:09.54 &-10:34:01.51 &32000& ...& 3.66& ...& 385& ...& H22\\
		&&& 32200 & 500 & 3.70 &0.06 & 374 & 11 & TW\\
		HD 189957 &20:01:00.00 &42:00:30.82 &32100& ...& 3.58& ...& 88& ...& H22\\
		&&& 32400 & 400 & 3.54 &0.06 & 100 & 15 & TW\\
		HD 191423 &20:08:07.11 &42:36:21.96 &32300& ...& 3.71& ...& 397& 18& H22; B23\\
		&&& 32400 & 600 & 3.54 &0.07 & 409 & 20 & TW\\
		HD 225160 &00:04:03.79 &62:13:19.01 &33200& ...& 3.35& ...& 77& ...& H22\\
		&&& 33300 & 800 & 3.35 &0.07 & 126 & 33 & TW\\
		HD 17603 &02:51:47.80 &57:02:54.47 &33300& ...& 3.26& ...& 103& ...& H22\\
		&&& 33300 & 900 & 3.34 &0.07 & 147 & 28 & TW\\
		HD 34078 &05:16:18.15 &34:18:44.34 &34500& ...& 4.07& ...& 13& ...& H22\\
		&&& 33300 & 500 & 4.07 &0.11 & 36 & 22 & TW\\
		HD 24431 &03:55:38.42 &52:38:28.75 &34900& ...& 3.77& ...& 49& ...& H22\\
		&&& 32900 & 800 & 3.77 &0.11 & 97 & 22 & TW\\
		HD 41161 &06:05:52.46 &48:14:57.42 &35200& ...& 3.84& ...& 331& ...& H22\\
		&&& 35700 & 900 & 3.99 &0.10 & 299 & 22 & TW\\
		HD 175876 &18:58:10.77 &-20:25:25.54 &36100& ...& 3.59& ...& 282& ...& H22\\
		&&& 37900 & 1600 & 3.75 &0.11 & 258 & 26 & TW\\
		HD 5689 &00:59:47.59 &63:36:28.25 &36800& ...& 3.72& ...& 255& ...& H22\\
		&&& 38000 & 1100 & 3.81 &0.08 & 223 & 17 & TW\\
		HD 36879 &05:35:40.53 &21:24:11.72 &36900& ...& 3.77& ...& 209& ...& H22\\
		&&& 36900 & 900 & 3.72 &0.06 & 199 & 18 & TW\\
		HD 42088 &06:09:39.57 &20:29:15.45 &40000& ...& 3.99& ...& 49& ...& H22\\
		&&& 41200 & 1200 & 4.12 &0.08 & 55 & 31 & TW\\
		BD+60 134 &00:56:14.21 &61:45:36.97 &40700& ...& 3.95& ...& 234& 9& H22; B23\\
		&&& 44800 & 1900 & 4.22 &0.13 & 219 & 23 & TW\\
	\enddata
	\tablecomments{Columns: (1) Common identifier for the comparison star. (2) Right Ascension of the comparison star in HH:MM:SS. (3) Declination of the comparison star in $^\circ$:':''. (4) Temperature of the comparison star in units of \kelvin. (5) Uncertaintity in the reported temperature in  units of \kelvin. (6) $\log$ surface-gravity. (7) Uncertainty in reported $\log$ surface-gravity. (8) Projected rotational velocity in units of \kms. (9) Uncertainty in reported projected rotational velocity in \kms. Values reported as ... indicate no available information in the literature. (10) The shortened reference for the literature stellar parameters values. In the case of two references, the first reference was used for \teff~and~\logg~and the second for \vsini. Reference code: D07---\cite{Daflon2007}. H10---\cite{Huang2010}. N12---\cite{Nieva2012}. M15---\cite{Mugnes2015}. S17---\cite{SimonDiaz2017}. H22---\cite{Holgado2022}. W22---\cite{Wessmayer2022}. B23---\cite{Britavskiy2023}. TW---This work.}
\end{deluxetable*}

Table~\ref{tab:comp} presents this work's analysis on the comparison star sample. Alternating rows list literature results and measurements from this work, respectively. Column 1 lists the common identifier of the comparison star. Columns 2 and 3 are the right ascension and declination in J2000 coordinates. Columns 4 and 5 report the temperature and its associated uncertainty. Columns 6 and 7 are the $\log$ surface gravity in cgs units and its uncertainty. Columns 8 and 9 are the projected rotational velocity and the uncertainty in this parameter. Column 10 is the reference for the measurement. For the purpose of comparison, we assume a temperature uncertainty of 1~k\kelvin, a gravity uncertainty of 0.10, and a rotational uncertainty of 10\% when none is listed. We sometimes used two works for full stellar parameter comparisons. In this case, the first reference listed in the table was used for temperature and gravity information, while the second was used for projected rotational velocity.

\begin{figure}[htb!]
        \plotone{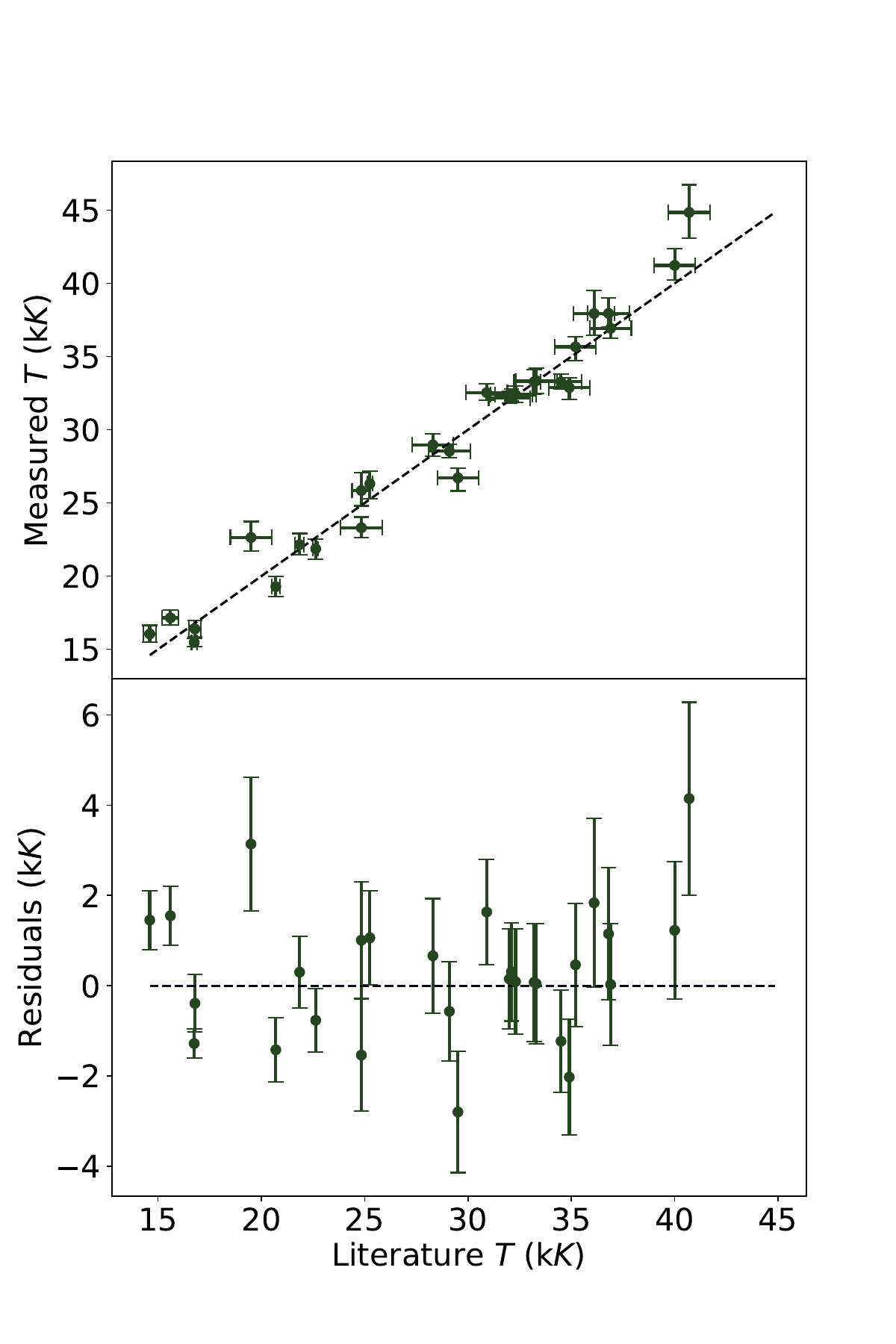}
        \caption{Temperature comparison between this work and values from the literature.}
        \label{fig:comp_temp}
\end{figure}

Figure~\ref{fig:comp_temp} plots our derived temperatures versus literature measurements (\textit{top panel}) and the residuals from the 1:1 relation (\textit{lower panel}). The error bars in the bottom panel reflect uncertainties in literature measurements and this work added in quadrature. Since this work reports both upper- and lower-uncertainties on each measurement, the maximal error is used in the residual uncertainty calculation. The dashed black horizontal line illustrates zero deviation. Our results on temperature for the comparison sample agree to within 1$\sigma$ of literature measurements across the entire sampled temperature range. Based on the residuals, we find our derived temperatures scatter about literature values by 1.5~k\kelvin.


\begin{figure}[htb!]
        \plotone{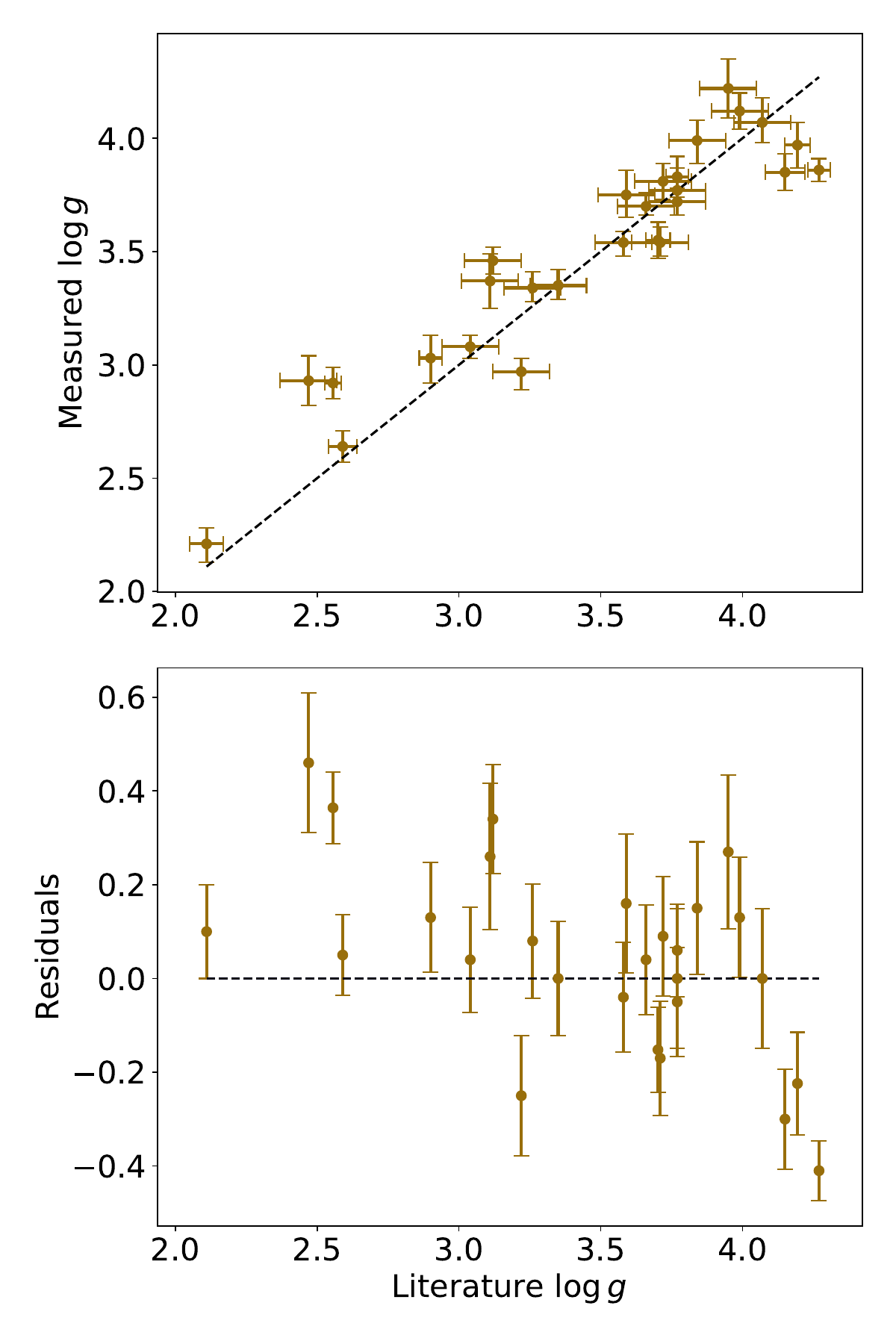}
        \caption{Gravity comparison between this work and values from the literature.}
        \label{fig:comp_g}
\end{figure}

Figure~\ref{fig:comp_g} plots our measured values for gravity against literature values (\textit{top panel}) and the residuals from the 1:1 relation (\textit{lower panel}). As in Figure~\ref{fig:comp_temp}, the error bars in the bottom panel reflect the quadrature sum of uncertainties from this work and those from the literature. Our results for gravity are generally in good agreement with literature measurements over the sampled gravity range with an RMS of 0.20.

\begin{figure}[htb!]
        \plotone{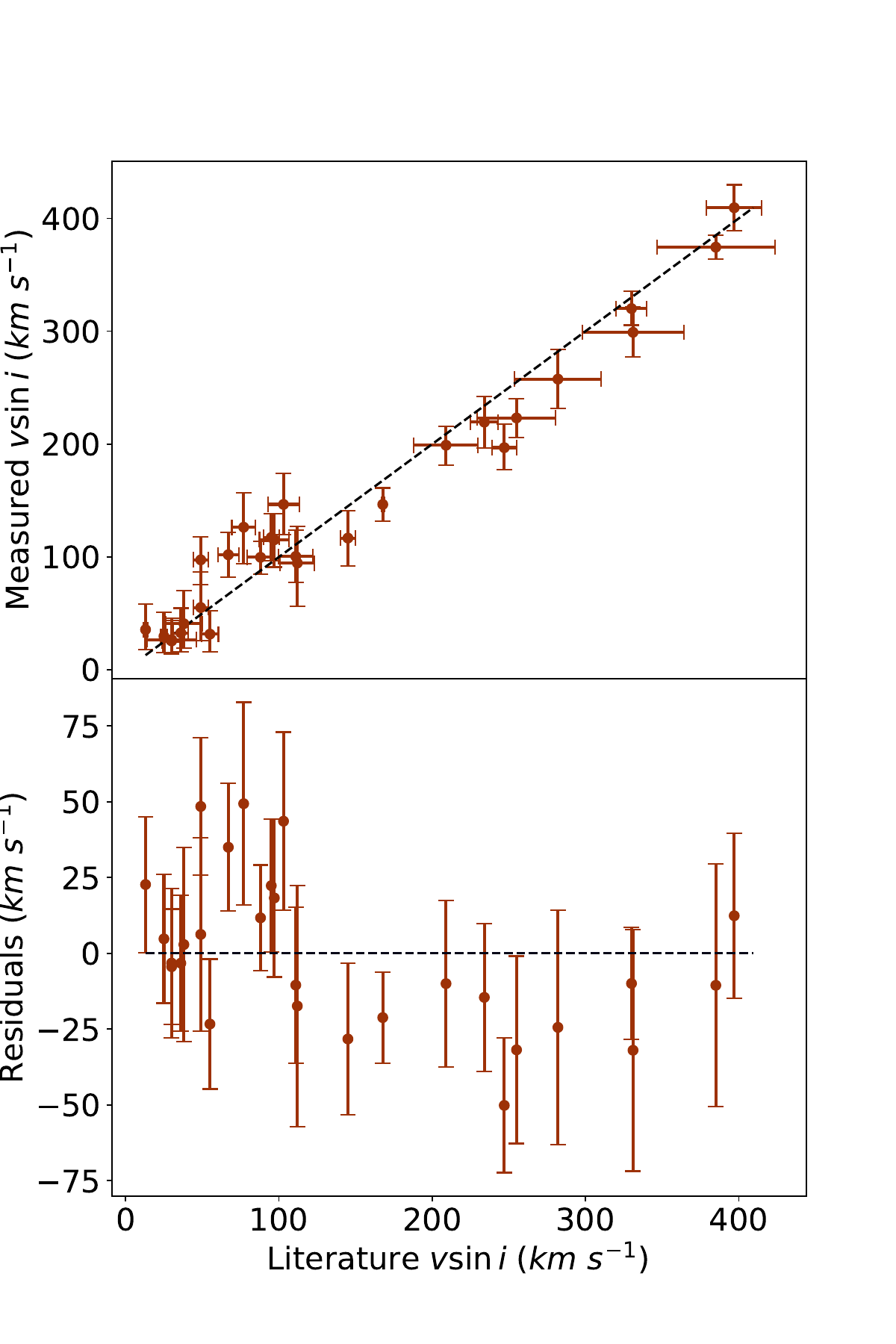}
        \caption{Projected rotational velocity comparison between this work and values from the literature.}
        \label{fig:comp_v}
\end{figure}

Figure~\ref{fig:comp_v} plots our rotational velocity against literature values (\textit{top panel}) and the residuals from the 1:1 relation (\textit{lower panel}). Error bars are the same as in the previous two figures. Residuals are large and positive below the resolution limit 100~\kms~and negative above this threshold. The instrumental resolution of KOSMOS of $R\sim$2200 corresponds to a velocity of approximately 100~\kms. Below the instrumental resolution, we conclude we cannot reliably measure projected rotation. Above this limit, our measurements appear to be systematically smaller than literature values by approximately 25~\kms. We cannot explain this deviation. However, it is noteworthy that measurements appear to agree within $1\sigma$, so we claim that our results serve as a reliable indicator of projected rotation, as our measurements are in agreement with the literature with an RMS of 25~\kms, and our data are capable of identifying stars with rotational velocities exceeding 200~\kms.


\subsubsection{Radii comparison}\label{sec:radiicomp}


From the results of the EXOFASTv2 analysis described in Section~\ref{sec:SED}, we obtained stellar radii and provide comparisons to results from the literature. For $\zeta$ Oph (BS013) we obtain a stellar radius of $8.46^{+0.47}_{-0.43}$~\rsun. Our measurement is in good agreement with the value of 8.5~\rsun~found in other works\footnote{Other works \citep{Herrero1992, HowarthandSmith2001, Shepard2022} have noted the oblate shape of $\zeta$ Oph due to its near-critical rotation and can measure distinctly the polar and equatorial radii. We cannot distinguish these radii, and our measured value of 8.46~\rsun~lies between measurements of polar radius of 7.0~\rsun~and measurements of equatorial radius of 9.1~\rsun.} \citep{Herrero1992, HowarthandSmith2001, Shepard2022}. For the comparison star HD 191423, we measure a stellar radius of $9.59^{+0.45}_{-0.43}$~\rsun, somewhat smaller than $18.3^{+8.7}_{-5.9}$~\rsun~as measured by \citet{Mahy2015}, obtained from SED modeling. For V* AE Aur, which is a member of both our comparison sample and our bowshock sample (BS375), we obtain stellar radius $6.44^{+0.28}_{-0.26}$~\rsun, in excellent agreement with the measurement of $6.8\pm0.5$~\rsun~from \citet{Aschenbrenner2023}.


\section{Properties of bowshock stars} \label{sec:results}

\begin{longrotatetable}
\begin{deluxetable}{llcccccccccccccl}
	\tablecaption{Bowshock Star Results\label{tab:res}}
	\tabletypesize{\scriptsize}
	\tablehead{
	ID & Sp. T. & $T_{\text{eff}}$ & $\sigma_{T_{\text{eff}}}$ & $\log g$ & $\sigma_{\log g}$ & \vsini & $\sigma_{v\sin i}$  & $R_*$ &$\sigma_{R_*}$ & $L_*$ & $\sigma_{L_*}$& $A_V$ & $\sigma_{A_V}$ & RUWE & flag\\
	 & & (\kelvin) &(\kelvin) & & & (\kms) & (\kms)& (\rsun) &(\rsun) & (\lsun) & (\lsun) & (mag) & (mag) &  &\\
	(1) &(2) &(3) &(4) &(5) &(6) &(7) &(8) &(9) &(10) & (11) & (12) & (13) & (14) & (15) &(16)}
	\startdata
		$001_\star$ & O+O e &34100 &2000 &3.83 &0.25 &216 &50 &10.9 &0.78 &134000 &24000 &3.52 &0.05 & 0.691 & SB2\\
		$007_\star$ & B2-3 V &22300 &1500 &4.44 &0.18 &58 &40 &3.57 &0.2 &2730 &660 &4.37 &0.14 & 0.873 & ...\\
		$010_\star$ & B1 III &27800 &1500 &3.81 &0.16 &43 &26 &8.03 &0.51 &27700 &5300 &2.61 &0.05 & 0.846 & ...\\
		$013_\star$ & O8-9 Vnn &32200 &500 &3.70 &0.06 &375 &11 &8.46 &0.47 &65500 &8500 &1.05 &0.06 & 4.489 & ...\\
		$016_\star$ & O+B &27400 &1900 &3.36 &0.18 &42 &26 &33.7 &9.2 &680000 &440000 &6.25 &0.13 & 0.892 & ...\\
		$019_\star$ & B2-3 III &24500 &2000 &3.48 &0.25 &47 &32 &35.4 &10.0 &520000 &340000 &6.26 &0.10 & 0.815 & ...\\
		$026_\star$ & B+B &28500 &1000 &3.90 &0.16 &417 &45 &15.29 &0.99 &144000 &24000 &5.14 &0.05 & 0.815 & SB2\\
		$027_\star$ & B5 III &22500 &1000 &3.96 &0.13 &61 &36 &5.46 &1.2 &7300 &4400 &3.28 &0.05 & 0.863 & ...\\
		$039_\star$ & O+B f &33300 &1300 &3.38 &0.13 &87 &50 &30.0 &6.7 &1010000 &540000 &6.07 &0.16 & 0.845 & SB2\\
		$043_\star$ & B2-3 I &26900 &1100 &3.74 &0.13 &32 &50 &8.25 &0.77 &33900 &7500 &3.77 &0.04 & 1.968 & R\\
		$051_\star$ & O+O &34000 &1000 &4.41 &0.21 &52 &58 &10.9 &1.7 &125000 &37000 &6.27 &0.07 & 1.032 & SB2\\
		$054_\star$ & B1 III(n) &27900 &1200 &3.01 &0.11 &138 &53 &18.3 &1.2 &125000 &19000 &2.67 &0.04 & 0.758 & SB1\\
		$056_\star$ & O+B &26800 &2000 &3.71 &0.24 &52 &35 &28.8 &6.0 &440000 &230000 &5.33 &0.07 & 1.864 & SB2; R\\
		$061_\star$ & B0 III(n) &30600 &600 &3.83 &0.12 &135 &19 &15.3 &1.4 &184000 &35000 &1.94 &0.03 & 1.415 & EB; R\\
		$063_\star$ & B+B &31100 &1100 &4.21 &0.21 &92 &33 &7.08 &0.42 &39100 &5200 &1.92 &0.04 & 0.988 & SB2\\
		$064_\star$ & O+B &46500 &2000 &4.15 &0.15 &234 &27 &9.19 &0.48 &345000 &59000 &4.18 &0.06 & 0.855 & SB2\\
		$065_\star$ & B5 III(n) &21200 &700 &3.72 &0.09 &91 &50 &6.22 &0.31 &6680 &960 &2.08 &0.07 & 0.869 & ...\\
		$066_\star$ & O+O &41300 &3000 &4.31 &0.25 &246 &50 &13.5 &4.3 &540000 &420000 &5.96 &0.19 & 1.023 & SB2\\
		$067_\star$ & O8-9 V &32200 &1200 &4.15 &0.21 &42 &27 &8.54 &0.5 &70300 &11000 &4.87 &0.08 & 1.002 & ...\\
		$150_\star$ & B1 V &25100 &1100 &3.97 &0.13 &49 &30 &8.42 &0.42 &27300 &4300 &5.25 &0.11 & 1.657 & R\\
		$200_\star$ & B2-3 I &25100 &1800 &3.50 &0.25 &48 &35 &32.8 &9.4 &370000 &260000 &6.16 &0.11 & 1.163 & ...\\
		$240_\star$ & B5 V &16500 &400 &3.55 &0.06 &87 &24 &5.97 &0.23 &2250 &260 &2.45 &0.05 & 0.855 & ...\\
		$255_\star$ & B2-3 V &20000 &800 &3.71 &0.10 &33 &20 &6.17 &0.29 &5290 &810 &3.44 &0.06 & 0.892 & ...\\
		$288_\star$ & B1 V(n) &27200 &1300 &3.88 &0.13 &169 &26 &8.79 &0.5 &42200 &7500 &2.86 &0.04 & 0.880 & ...\\
		$298_\star$ & B1 1 &22400 &1500 &2.98 &0.14 &56 &36 &17.7 &1.2 &65000 &16000 &2.04 &0.05 & 1.033 & ...\\
		$300_\star$ & B1 V &29200 &800 &4.06 &0.12 &74 &41 &6.64 &0.28 &29700 &3500 &2.73 &0.05 & 0.835 & ...\\
		$302_\star$ & B2-3 V(n) &22000 &1000 &3.79 &0.14 &211 &40 &8.54 &0.67 &16700 &3600 &2.56 &0.07 & 1.619 & R\\
		$303_\star$ & B2-3 III &20200 &1600 &2.70 &0.18 &45 &31 &68.4 &22.0 &790000 &570000 &5.99 &0.19 & 0.987 & SB1\\
		$305_\star$ & O+O &33500 &1000 &4.59 &0.15 &269 &45 &10.2 &1.5 &119000 &41000 &3.12 &0.11 & 5.203 & SB1; SB2; R\\
		$306_\star$ & B+B &27200 &3000 &3.58 &0.23 &218 &55 &9.61 &0.7 &53000 &17000 &2.86 &0.08 & 0.970 & SB2\\
		$314_\star$ & O+B &26500 &1400 &3.69 &0.15 &210 &28 &12.4 &1.5 &71000 &24000 &2.17 &0.11 & 0.987 & SB2\\
		$319_\star$ & B+B &23700 &1300 &3.08 &0.15 &47 &30 &35.6 &4.1 &424000 &120000 &4.17 &0.05 & 0.957 & SB1; SB2\\
		$320_\star$ & B1 I &27800 &1100 &3.09 &0.13 &37 &24 &23.9 &2.4 &300000 &79000 &2.31 &0.19 & 0.866 & ...\\
		$321_\star$ & B0 V(n) &30900 &1800 &4.24 &0.24 &126 &36 &9.36 &0.74 &79000 &17000 &4.64 &0.12 & 1.031 & ...\\
		$322_\star$ & O8-9 V &32700 &1000 &3.98 &0.17 &100 &24 &13.96 &0.95 &206000 &34000 &5.03 &0.08 & 1.045 & SB1\\
		$324_\star$ & O+O &32400 &1000 &4.01 &0.15 &207 &26 &17.3 &1.3 &306000 &54000 &3.39 &0.05 & 0.893 & SB2\\
		$325_\star$ & B2-3 III &21600 &800 &3.29 &0.10 &33 &20 &15.35 &0.82 &47100 &7600 &2.81 &0.06 & 1.556 & R\\
		$326_\star$ & B1 I &25300 &1700 &3.71 &0.29 &61 &45 &21.0 &1.4 &196000 &45000 &4.33 &0.06 & 1.077 & ...\\
		$328_\star$ & B+B &26000 &1500 &3.97 &0.23 &158 &62 &9.6 &0.61 &390000 &7000 &3.96 &0.06 & 0.923 & SB2\\
		$330_\star$ & B1 I &22400 &1000 &2.94 &0.12 &42 &27 &28.4 &1.4 &177000 &28000 &3.60 &0.05 & 1.041 & ...\\
		$331_\star$ & O+O &34700 &1000 &3.92 &0.13 &192 &20 &16.9 &0.43 &68800 &7800 &5.58 &0.04 & 1.080 & SB1; SB2\\
		$332_\star$ & O+O &32200 &800 &3.95 &0.13 &199 &22 &8.64 &0.43 &68800 &7800 &5.58 &0.04 & 1.051 & SB1; SB2\\
		$335_\star$ & B1 I &24400 &1200 &3.32 &0.14 &52 &29 &16.99 &0.77 &82500 &9500 &4.87 &0.09 & 1.056 & ...\\
		$339_\star$ & O+O &45600 &4500 &4.28 &0.33 &200 &75 &16.1 &5.0 &990000 &830000 &6.29 &0.12 & 1.031 & SB2\\
		$340_\star$ & B1 I &25800 &1200 &3.42 &0.12 &29 &20 &11.24 &0.56 &48000 &7800 &1.26 &0.05 & 0.942 & ...\\
		$341_\star$ & O+O &33100 &800 &4.32 &0.16 &150 &23 &13.4 &2.8 &196000 &93000 &6.16 &0.09 & 1.301 & SB1; SB2\\
		$343_\star$ & O+B &23600 &1500 &2.57 &0.18 &96 &50 &54.3 &3.0 &820000 &140000 &3.55 &0.04 & 0.897 & SB2\\
		$344_\star$ & O6 If+ &46800 &900 &3.78 &0.05 &154 &60 &21.69 &0.87 &2030000 &210000 &5.60 &0.07 & 0.937 & ...\\
		$345_\star$ & B0 III &27400 &900 &3.74 &0.09 &127 &24 &7.38 &0.29 &26500 &3300 &2.53 &0.06 & 0.874 & ...\\
		$346_\star$ & B1 III &21400 &1100 &3.19 &0.14 &96 &40 &40.0 &21.0 &380000 &310000 &6.26 &0.61 & 1.304 & SB1\\
		$349_\star$ & O+B &25200 &1300 &3.28 &0.12 &235 &20 &13.03 &0.82 &559000 &11000 &2.65 &0.05 & 0.916 & SB2; R\\
		$350_\star$ & B1 III &24200 &1600 &3.16 &0.17 &91 &40 &35.8 &8.2 &450000 &270000 &5.14 &0.07 & 3.800 & SB1; R\\
		$351_\star$ & O8-9 III(n) &33100 &700 &4.03 &0.12 &190 &20 &16.6 &2.2 &305000 &87000 &2.26 &0.02 & 1.058 & ...\\
		$353_\star$ & O+B &37700 &1300 &3.98 &0.12 &250 &25 &14.79 &0.99 &408000 &68000 &4.52 &0.06 & 0.881 & SB2\\
		$354_\star$ & O+B e &24800 &2400 &2.93 &0.21 &448 &74 &38.5 &5.0 &530000 &220000 &4.76 &0.20 & 0.996 & SB2\\
		$356_\star$ & B1 III &26200 &1000 &3.25 &0.10 &137 &30 &13.75 &0.69 &72700 &11000 &2.66 &0.04 & 0.986 & SB1\\
		$357_\star$ & B2-3 III &24200 &1000 &3.24 &0.10 &113 &30 &18.1 &2.0 &98000 &28000 &2.74 &0.04 & 3.035 & SB1; R\\
		$358_\star$ & B5 III &20500 &700 &3.83 &0.08 &45 &30 &4.27 &0.16 &2620 &320 &2.52 &0.06 & 0.981 & ...\\
		$359_\star$ & B0 V &31700 &400 &3.59 &0.06 &54 &27 &13.12 &0.62 &156000 &16000 &2.26 &0.02 & 0.974 & ...\\
		$360_\star$ & B1 I &21500 &800 &2.58 &0.09 &41 &10 &65.9 &4.3 &840000 &140000 &2.79 &0.11 & 0.989 & SB1\\
		$361_\star$ & B2-3 I &23200 &1800 &2.69 &0.17 &51 &20 &44.3 &3.0 &609000 &100000 &4.62 &0.03 & 1.046 & SB1\\
		$362_\star$ & B+B &25100 &2100 &3.34 &0.21 &237 &38 &15.2 &1.1 &105000 &27000 &5.54 &0.06 & 1.153 & SB2\\
		$364_\star$ & O7 V &34600 &600 &4.37 &0.12 &77 &24 &9.06 &0.72 &106000 &19000 &1.87 &0.03 & 2.220 & R\\
		$365_\star$ & B1 III &26800 &1200 &3.54 &0.13 &29 &20 &9.29 &0.45 &37900 &6200 &2.24 &0.08 & 0.913 & ...\\
		$366_\star$ & B0 V &30200 &400 &4.02 &0.05 &89 &20 &18.04 &1.0 &248000 &32000 &0.89 &0.04 & 1.014 & ...\\
		$367_\star$ & O8-9 V &32500 &900 &3.95 &0.15 &52 &30 &11.6 &0.54 &138000 &17000 &4.25 &0.04 & 1.140 & ...\\
		$368_\star$ & O+O &32300 &1000 &3.91 &0.17 &308 &30 &9.26 &0.5 &84300 &11000 &4.17 &0.04 & 1.246 & SB2\\
		$369_\star$ & O7 V &37300 &1000 &4.22 &0.10 &89 &26 &11.18 &1.2 &221000 &52000 &1.96 &0.05 & 2.736 & R\\
		$371_\star$ & O5 V &44300 &2500 &4.23 &0.18 &120 &46 &8.97 &0.69 &272000 &69000 &3.96 &0.04 & 1.294 & SB1\\
		$372_\star$ & O8-9 V(n) &31900 &1000 &3.93 &0.15 &142 &26 &12.22 &0.87 &143000 &24000 &3.84 &0.07 & 1.107 & SB1\\
		$373_\star$ & B5 III &23700 &500 &3.73 &0.06 &33 &21 &5.94 &0.28 &9400 &1200 &0.58 &0.04 & 1.178 & ...\\
		$374_\star$ & O+B &33300 &500 &3.56 &0.06 &290 &16 &12.7 &1.2 &172000 &38000 &0.76 &0.04 & 1.217 & SB2\\
		$375_\star$ & O8-9 III &33200 &500 &4.06 &0.09 &38 &23 &6.44 &0.28 &44600 &4400 &1.80 &0.06 & 0.993 & ...\\
		$376_\star$ & O+B &43500 &1600 &4.04 &0.11 &185 &27 &11.03 &0.75 &386000 &75000 &1.00 &0.10 & 0.949 & SB2\\
		$377_\star$ & O+B &37600 &1100 &3.97 &0.09 &104 &29 &9.93 &0.46 &170000 &22000 &1.97 &0.02 & 0.930 & SB2\\
		$378_\star$ & O8-9 V &32700 &500 &4.19 &0.08 &39 &21 &5.94 &0.42 &36700 &5500 &0.14 &0.08 & 1.179 & ...\\
		$381_\star$ & O6 V &39900 &1200 &4.03 &0.10 &107 &23 &13.9 &1.3 &450000 &95000 &1.34 &0.13 & 1.105 & ...\\
		$383_\star$ & O+B f &34300 &600 &3.31 &0.05 &145 &43 &26.9 &2.2 &890000 &170000 &2.33 &0.04 & 1.033 & SB2\\
		$384_\star$ & O+B &35400 &1500 &4.26 &0.22 &246 &42 &12.53 &0.77 &229000 &37000 &3.10 &0.06 & 0.973 & SB2\\
		$385_\star$ & O+B (f) &40300 &2100 &3.81 &0.11 &227 &22 &12.68 &0.67 &356000 &69000 &3.33 &0.03 & 0.985 & SB2\\
		$386_\star$ & B1 I &24500 &1100 &2.80 &0.10 &38 &26 &38.3 &2.1 &439000 &82000 &2.35 &0.07 & 0.880 & EB\\
		$388_\star$ & B2-3 I &21800 &1100 &2.71 &0.11 &32 &23 &57.3 &6.1 &630000 &180000 &3.40 &0.06 & 2.496 & ...\\
		$389_\star$ & B2-3 I &24400 &1000 &3.31 &0.10 &28 &19 &15.65 &1.0 &71000 &13000 &2.39 &0.04 & 1.383 & ...\\
		$634_\star$ & O8-9 V(n) &31700 &700 &4.18 &0.10 &149 &19 &11.2 &0.51 &116000 &13000 &2.78 &0.05 & 0.829 & ...\\
		$640_\star$ & O+B &27300 &1500 &4.28 &0.16 &256 &38 &10.13 &0.65 &54400 &11000 &3.37 &0.08 & 1.408 & SB2; R\\
		$662_\star$ & O8-9 III(n) &32800 &900 &3.81 &0.13 &154 &19 &13.42 &0.8 &187000 &27000 &1.56 &0.07 & 0.816 & ...\\
		$667_\star$ & B0 III(n) &29600 &700 &4.05 &0.10 &153 &16 &10.05 &1.1 &70000 &15000 &1.23 &0.16 & 0.825 & EB\\
		$673_\star$ & B1 V &27100 &1800 &4.01 &0.29 &56 &45 &9.1 &1.8 &37500 &18000 &5.49 &0.37 & 1.375 & ...\\
		$687_\star$ & B+B &27100 &2100 &4.19 &0.35 &261 &130 &19.6 &9.9 &190000 &210000 &6.31 &0.28 & 1.028 & SB2\\
		$692_\star$ & O+O &36800 &1100 &4.60 &0.14 &110 &51 &10.62 &0.52 &190000 &24000 &5.21 &0.03 & 1.055 & SB2\\
		$700_\star$ & B+B &27000 &1900 &3.82 &0.25 &158 &79 &25.0 &11.0 &420000 &340000 &6.29 &0.37 & 0.884 & SB2\\
		$705_\star$ & B1 III &27100 &1500 &3.45 &0.16 &132 &51 &14.8 &0.9 &109000 &22000 &3.99 &0.07 & 0.944 & ...\\
		2G0063185-0066622$_\star$ & B+B &24000 &1600 &3.04 &0.19 &46 &36 &43.0 &11.0 &58000 &380000 &6.23 &0.13 & 0.874 & SB2\\
		2G0065094-0061344$_\star$ & B+B &27900 &1700 &3.71 &0.18 &217 &35 &7.35 &0.49 &28100 &6500 &4.73 &0.07 & 1.303 & SB2\\
		2G0118310-0062990$_\star$ & B2-3 III &23000 &800 &3.43 &0.10 &30 &21 &11.51 &0.59 &31400 &4900 &1.70 &0.05 & 0.885 & ...\\
		2G0592097+0012519$_\star$ & B+B &26200 &2000 &3.69 &0.24 &190 &54 &29.5 &10.0 &420000 &350000 &6.36 &0.15 & 2.037 & SB2; R\\
		2G0600615-0018340$_\star$ & B0 V &29800 &700 &3.96 &0.12 &151 &17 &8.18 &0.37 &49000 &5200 &3.00 &0.04 & 0.847 & ...\\
		2G0770694+0193568$_\star$ & B0 V &30100 &600 &4.03 &0.09 &103 &20 &7.24 &0.55 &39400 &6700 &2.27 &0.04 & 3.412 & R\\
		2G1045666+0128085$_\star$ & O+B &34900 &500 &4.64 &0.11 &243 &30 &11.42 &0.51 &176000 &51000 &3.01 &0.05 & 1.069 & SB2\\
		$\kappa$ Cas & B1 I &22800 &1200 &2.72 &0.11 &33 &23 &81.1 &7.4 &1740000 &430000 &1.13 &0.10 & 2.368 & ...\\
		2G3525610+0001887$_\star$ & B+B &26700 &1700 &3.77 &0.19 &129 &54 &10.58 &0.71 &56000 &13000 &4.75 &0.10 & 1.021 & SB2\\
		2G3527642+0032660$_\star$ & F &-99 &-99 &-99 &-99 &-99 &-99 &-99 &-99 &-99 &-99 &-99 &-99 & 0.923 & ...\\
		2G3533026+0008095$_\star$ & O+B &39600 &1800 &3.97 &0.14 &209 &22 &13.86 &0.73 &435000 &73000 &4.48 &0.04 & 0.921 & SB2\\
		2G3585210+0089852$_\star$ & O+B &34700 &500 &4.44 &0.13 &181 &19 &11.14 &0.75 &164000 &24000 &5.08 &0.07 & 0.930 & SB2\\
	\enddata
	\tablecomments{Columns: (1) Stellar bowshock nebuae identifier from \citet{Kobulnicky2016} and addendum \citet{Jayasinghe2019} catalogs; (2) Approximate MK spectral type; (3) Measured \teff~in \kelvin; (4) Uncertainty on \teff in \kelvin; (5) Measured \logg; (6) Uncertainty on \logg; (7) Measured \vsini~in \kms; (8) Uncertainty on \vsini ~in \kms; (9) Stellar radius in \rsun; (10) Uncertainty on stellar radius in \rsun; (11) Luminosity in \lsun; (12) Uncertainty on luminosity in \lsun; (13) Visual-band extinction in magnitudes; (14) Uncertainty on $A_V$ in magnitudes. (15) Gaia RUWE value (16) Binary flag. Key: SB1 - Probability of single star $<0.05$, as found in \citet{Chick2020}; SB2 - Double-line spectroscopic binary; EB- Eclipsing binary in \citet{Malkov2006}; R - Gaia RUWE $> 1.4$; ... - no apparent indicators of binarity.}
\end{deluxetable}
\end{longrotatetable}

Table~\ref{tab:res} lists the results of the analysis on the program stars. Column~1 gives the bowshock star identifier from the \citet{Kobulnicky2016} and \citet{Jayasinghe2019} catalogs. Column~2 gives the approximate MK spectral classification. We use the criteria developed in \citet{Liu2019} to determine the approximate spectral type, with further qualifiers explained in \citet{Sota2011}. For spectroscopic binaries, we report a classification of two O stars, an O and a B star, or two B stars, based on the presence and/or absence of \ion{He}{1} and \ion{He}{2} lines.  Columns~3 and 4 give the measured temperature and uncertainty in temperature from spectra. Columns~5 and 6 list the measured \logg~and uncertainty from spectra. Columns~7 and 8 report the measured projected rotational velocity and its uncertainty, also measured from spectra. Typical uncertainties are $1000$~\kelvin~ in temperature, $0.10$ in \logg, and 40~\kms~in \vsini. Columns~9 and 10 report the stellar radius from EXOFASTv2, and its uncertainty. Columns~11 and 12 give the luminosity and uncertainty measured from EXOFASTv2. Lastly, columns~13 and 14 give the extinction in V band, $A_V$, and uncertainty, as measured in EXOFASTv2. For dubious bowshock stars $2G3527642+0032660_\star$, we exclude this star from analysis presented in this paper, and list $-99$ to indicate null results on stellar parameters. See Section~\ref{app:HD157642} in the Appendix for more discussion on this star.

\subsection{Temperature} \label{sec:Tresults}

\begin{figure}[htb!]
        \plotone{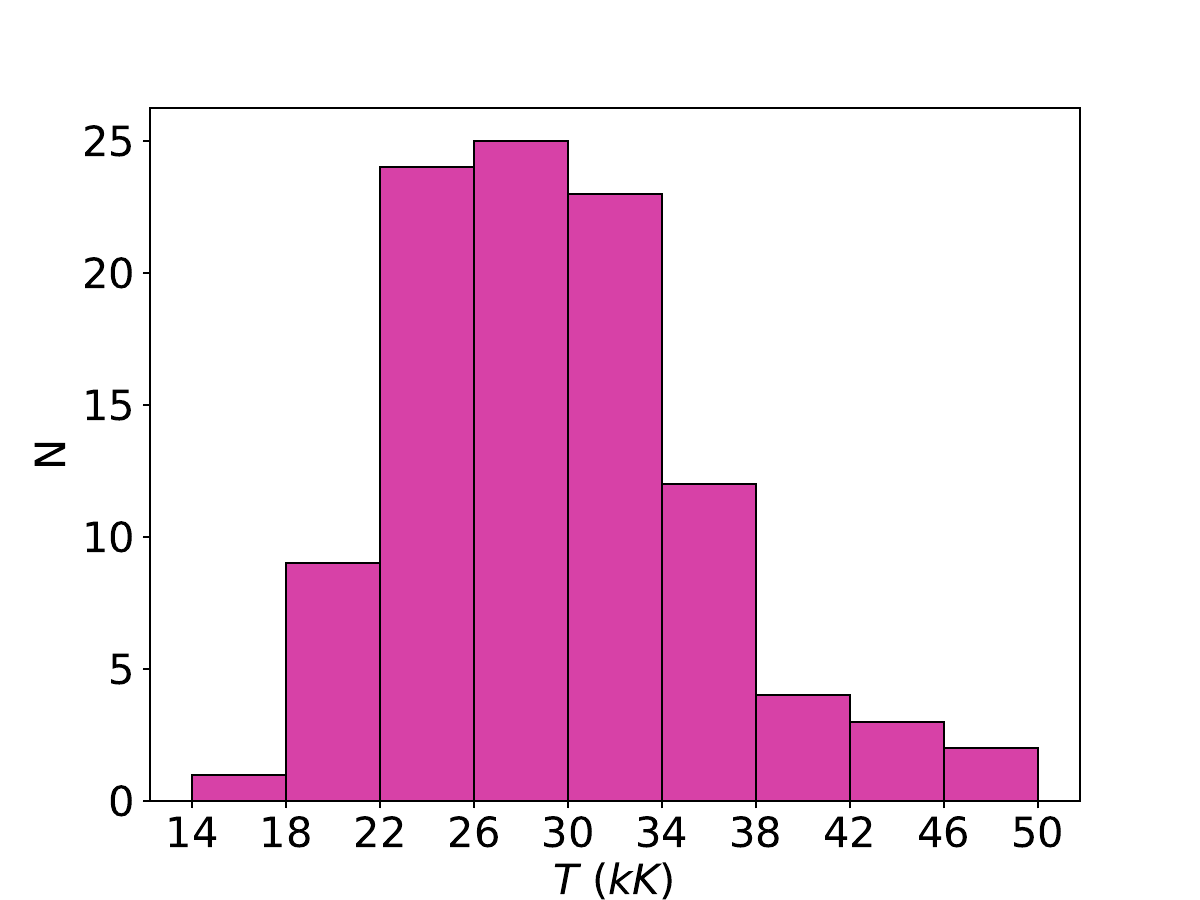}
        \caption{Temperature distribution of bowshock stars.}
        \label{fig:T_hist}
\end{figure}

Figure~\ref{fig:T_hist} presents a histogram of retrieved temperatures for the bowshock stars in bins of 4000~\kelvin. We find temperatures ranging from 16500~\kelvin~to 46800~\kelvin, with a median temperatures 27800~\kelvin. Defining a criterion of B stars having effective temperatures lower than 30~k\kelvin, we find our sample of 103 OB program stars consists of \numBstars~B stars (\fracB) and \numOstars~O stars (\fracO). However, this should be not be interpreted as a true characterization of the parent sample given our magnitude-limited selection criteria.

\subsection{Surface gravity} \label{sec:gresults}

\begin{figure}[htb!]
        \plotone{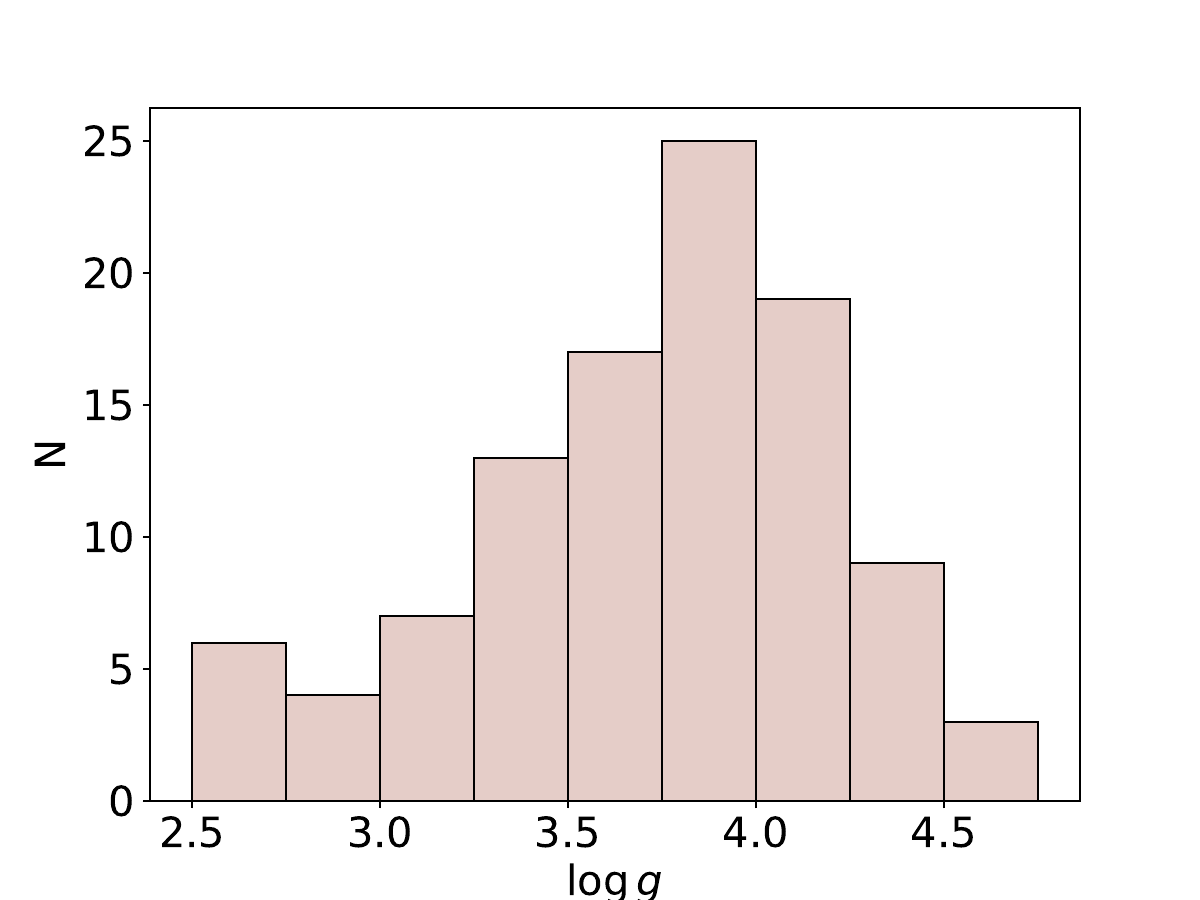}
        \caption{\logg~ distribution of bowshock stars.}
        \label{fig:g_hist}
\end{figure}

Figure~\ref{fig:g_hist} presents the distribution of \logg~ among the bowshock stars in bins of 0.25. We find $\log$ surface gravities ranging from 2.57 to 4.60, with a median value of 3.81. This is consistent with O- and B-type stars ranging in evolutionary stages dwarf to supergiant.

\subsection{Projected rotational velocity}

\begin{figure}[htb!]
        \plotone{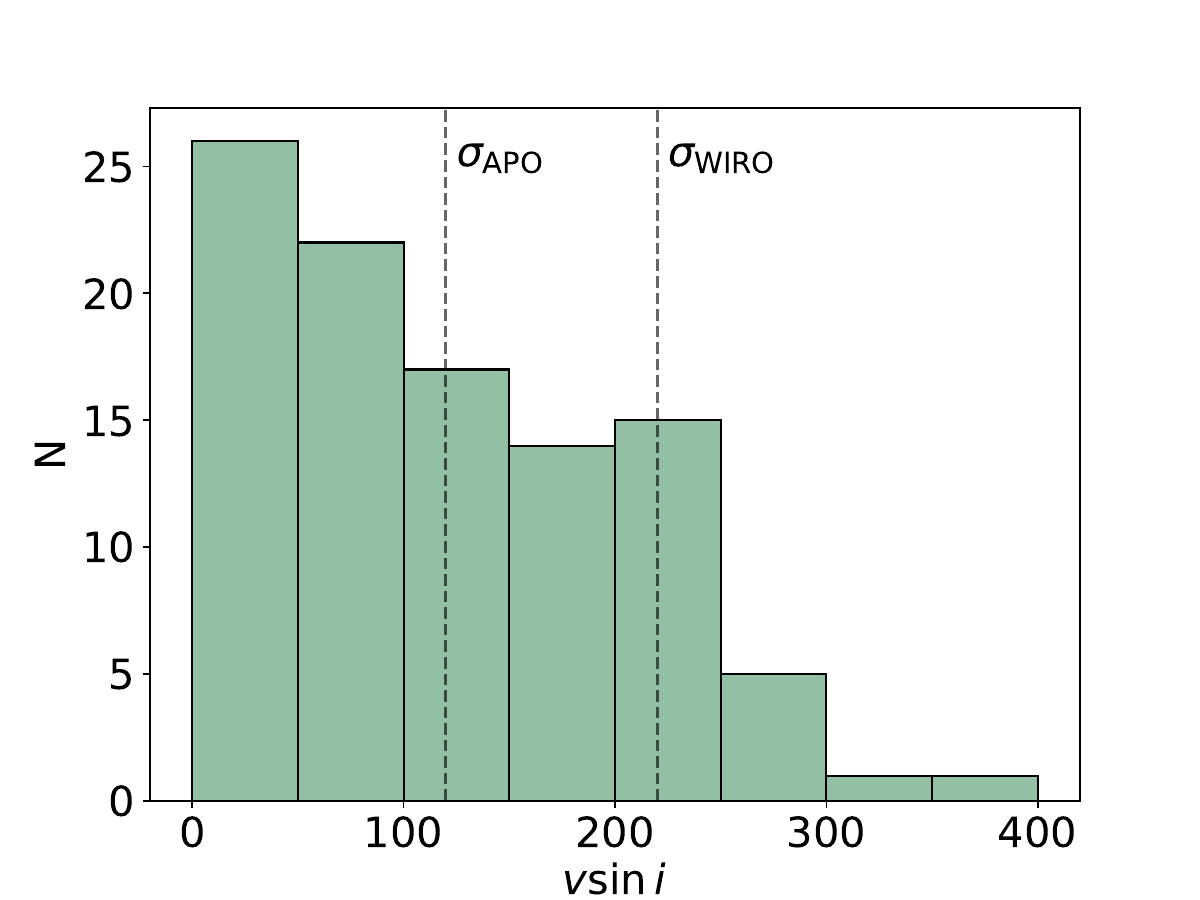}
        \caption{Projected rotational velocity distribution of bowshock stars.}
        \label{fig:v_hist}
\end{figure}

Figure~\ref{fig:v_hist} presents the distribution of projected rotational velocity for the program stars in bins of 50~\kms. Due to the limitations of low-resolution spectrographs described in Section~\ref{sec:specconfig}, a broadening of 100~\kms~should be regarded as a floor, below which we cannot be certain of the projected rotational velocity. We find projected rotational velocities ranging from 18 to 375~\kms, with a median value of 100~\kms. A majority (\fracslow) of program stars can be described as ``slow rotators,'' with a projected rotational velocity below 120~\kms.

\subsection{Binarity}

We find 62 bowshock stars (60\% of the sample) show some indications of binarity. Of the total \numstars bowshock stars, we find 16 (15\%) are single-lined spectroscopic binaries, 41 (39\%) are double-lined spectroscopic binaries, 3 (3\%) are eclipsing binaries, and 15 (14\%) have Gaia RUWE $>$1.4. Several bowshock stars show several indications of binarity, hence why the percentages add up to more than 100\%. In the case of composite spectra (i.e. SB2) we expect the fitted spectral parameters represent some average of both components of the system with larger uncertainties, in which a single star model doesn't fit well. Our analysis is most problematic for de-blended SB2's, in which \vsini\ is erroneously inflated to cover split lines.






\subsection{Visual extinction}

\citet{Kobulnicky2016} presented $K$-band extinctions $A_K$ for their sample of 709 stellar bowshock candidates based on $H-[4.5]$ color excess method described in \citet{Majewski2011}. Using the \citet{Cardelli1989} extinction curve for average Galactic dust $R_V=$3.1, we compute the corresponding $A_V$ using the relation $A_V = 8.13A_K$.

\begin{figure}[htb!]
        \plotone{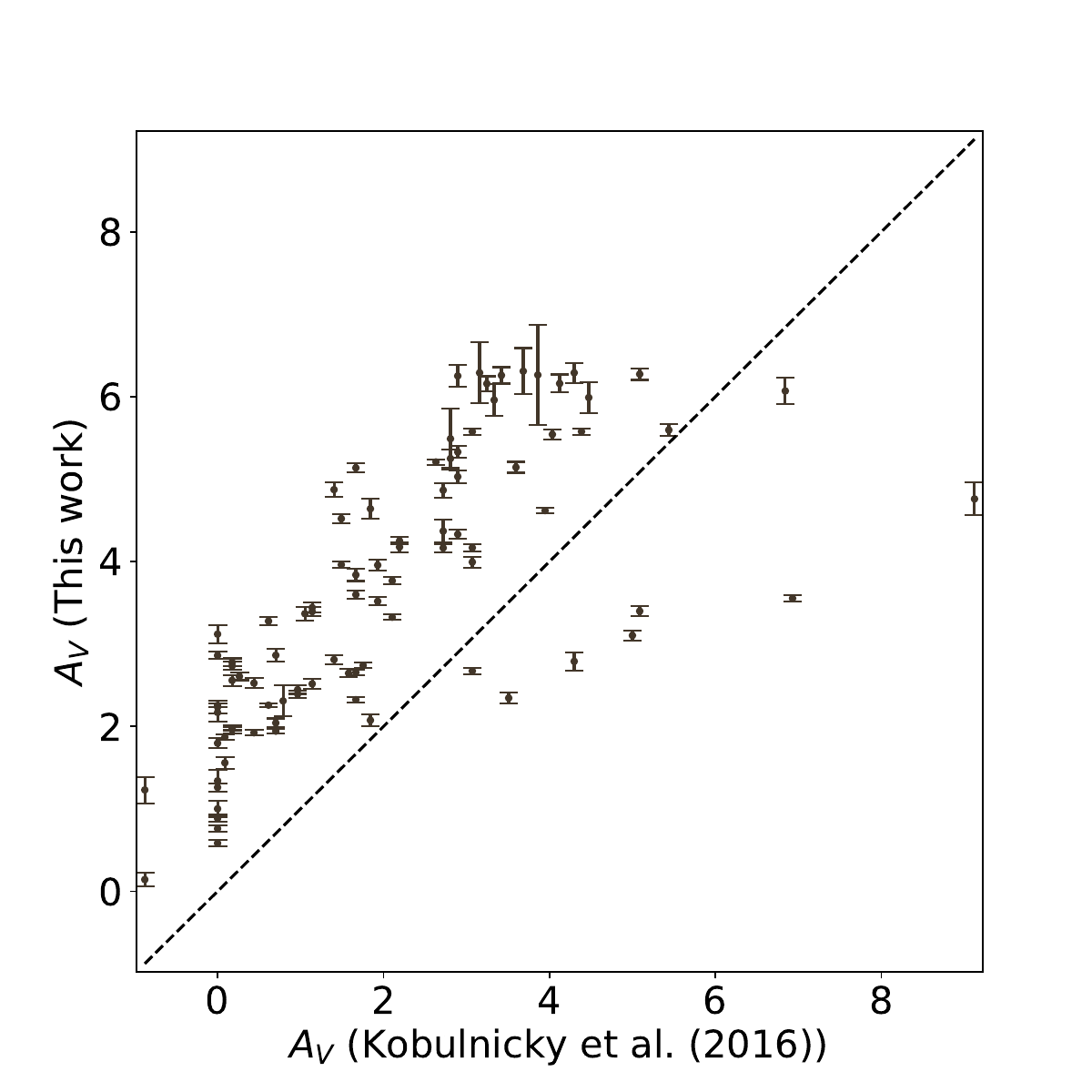}
        \caption{Visual-band extinction compared with transformed extinctions from previous work.}
        \label{fig:extinc}
\end{figure}


Figure~\ref{fig:extinc} compares extinctions derived in this work using EXOFASTv2 (\textit{vertical axis}) with transformed $K$-band extinctions from Rayleigh-Jeans Color Excess (RJCE) methods (\textit{horizontal axis}). Our extinction derived from SED fitting are systematically higher than the RJCE extinctions in \citet{Kobulnicky2016}. While there is a strong correlation between the two methods, the SED-fitting derived extinctions are 2--3 mags larger than the RJCE values. The systematic difference between these two methods of measuring extinction indicates that $H-[4.5]$~\um~ color-excess technique is inappropriate for early-type stars, as noted by \citet{Majewski2011}. The RJCE method described in \citet{Majewski2011} infers a $K-$band extinction from a $H-[4.5]$~\um~ color. For spectral types A--M, \citet{Majewski2011} demonstrate that this color only varies by 0.1~mag. With the inclusion of B-type stars, this color varies by 0.4~mag. With the inclusion of O spectral types, this color trend may diverge as the atmospheres of stars become increasingly less affected by absorption from metals in the infrared spectrum with increasing temperature. We conclude that the SED-fitting techniques provide a more reliable measure of extinction than RJCE methods for early-type stars.

\section{Discussion} \label{sec:discussion}

\subsection{Are bowshock stars distinct from typical OB stars?}

\begin{figure*}[htb!]
        \plotone{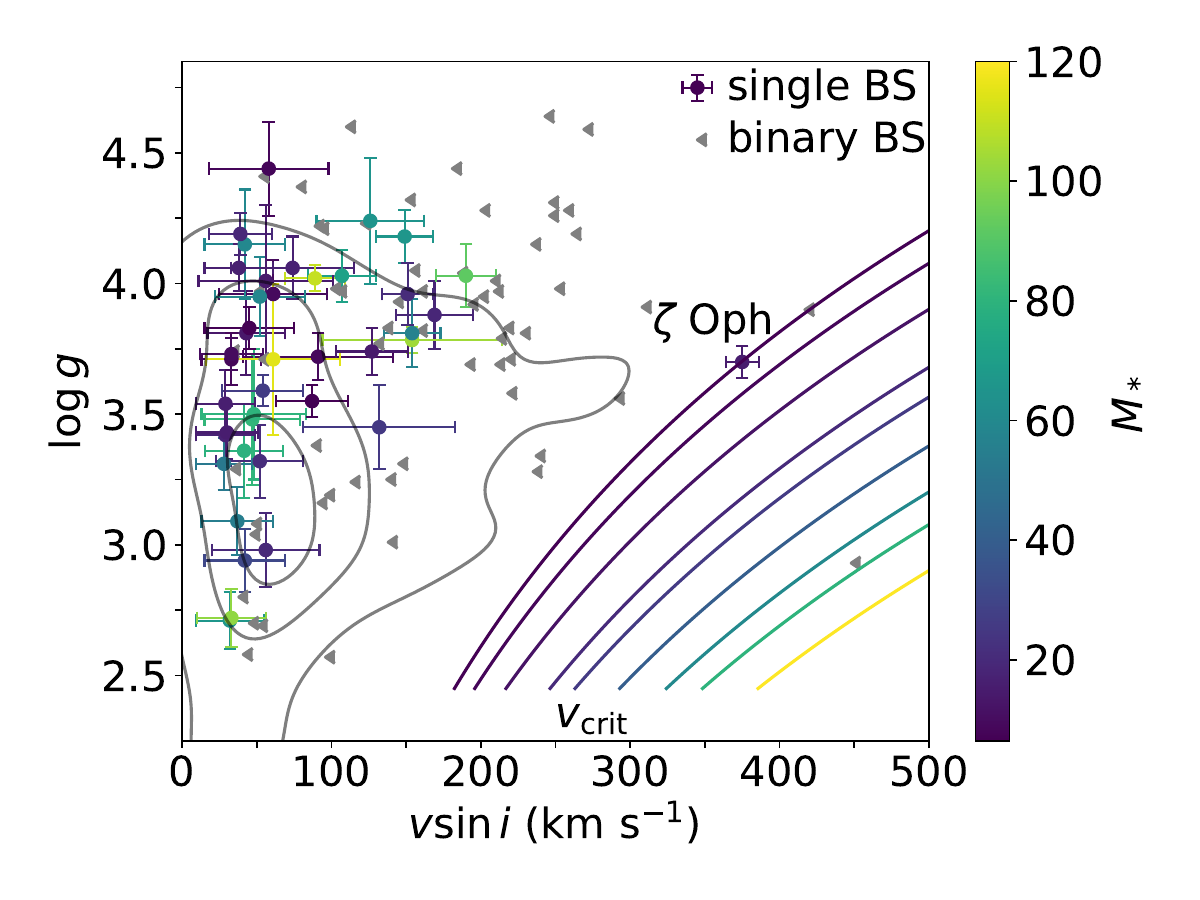}
        \caption{\logg~versus \vsini~for single bowshock stars (\textit{colored points}), binary bowshock stars (\textit{grey triangles}), and 16/50/84th percentile levels for stars drawn from the IACOB sample. The colored tracks show theoretical maximum rotational velocities for a range of masses indicated by the color bar, with an Eddington factor $\Gamma=0.30$ (\textit{solid colored tracks}).}
        \label{fig:g_vsini}
\end{figure*}

Figure~\ref{fig:g_vsini} plots \logg~versus \vsini~for our sample of bowshock stars with no apparent indications of binarity (\textit{filled circles colored by stellar mass}) and probable binary bowshock stars (\textit{grey triangles}). For binary bowshock stars, we consider our measured \vsini\ to be an upper limit. Overlaid in the gray contours are the 16, 50, and 84th percentile levels drawn from a sample of 285 Galactic O stars combined with 527 late-O and early-to-mid-B supergiants studied in the IACOB project \citep{Holgado2022, deBurgos2024}. Colored tracks indicate maximum rotation rates, computed from the equation,

\begin{equation}
    v_{\text{crit}} = \sqrt{\frac{GM_*}{R_*}\left(1-\Gamma\right)}\text{ , }\label{eq: crit_rotation}
\end{equation}

\noindent taking into account gravitational, Thompson scattering, and centrifugal accelerations, as derived in seminal works \citep{Langer1997, Langer1998, Glatzel1998}. The critical rotation relation in Equation~\ref{eq: crit_rotation} bears a resemblance to the escape velocity, and the two are related by $v_{\text{crit}}=v_{\rm{esc}}/\sqrt{2}$. We choose Eddington $\Gamma$ of 0.30, a value typical for OB stars. \citep[][]{Verhamme2024}. $\zeta$ Oph, a rapid rotator, lies to the right of the figure near its critical rotation track with a projected rotational velocity of 375~\kms. We measure a stellar mass of 15~\msun, corresponding to an estimated critical velocity of 450~\kms. Other works find a similar critical velocity of $\sim$500~\kms~\citep{Shepard2022}. We find bowshock stars occupy a similar locus of \logg~and \vsini~as Galactic OB populations. With the exception of $\zeta$ Oph, the bowshock sample does not contain any stars close to the critical rotational velocity. A handful of bowshock stars lie above $\log g\geq$4.40 in a region not shared by \texttt{IACOB} stars, but these are almost exclusively probable binaries. The four bowshock stars displaced to the right of the panel are also probable binaries, with high measured rotational velocities due to the increased line width from the blended motions of two stars, which compromises the \vsini~measurement.











\begin{figure*}[htb!]
        \gridline{\fig{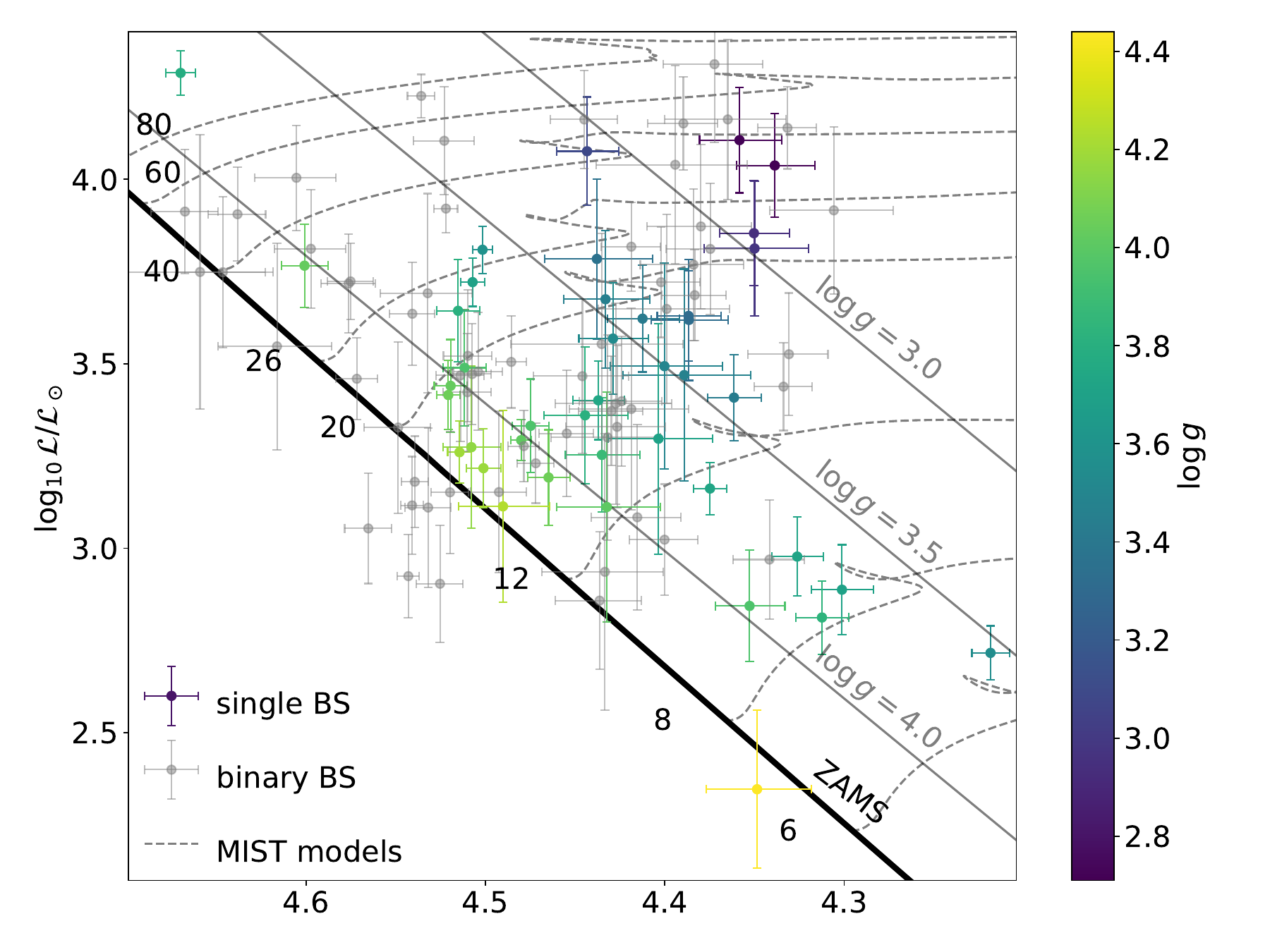}{0.72\textwidth}{(a)}}
        \gridline{\fig{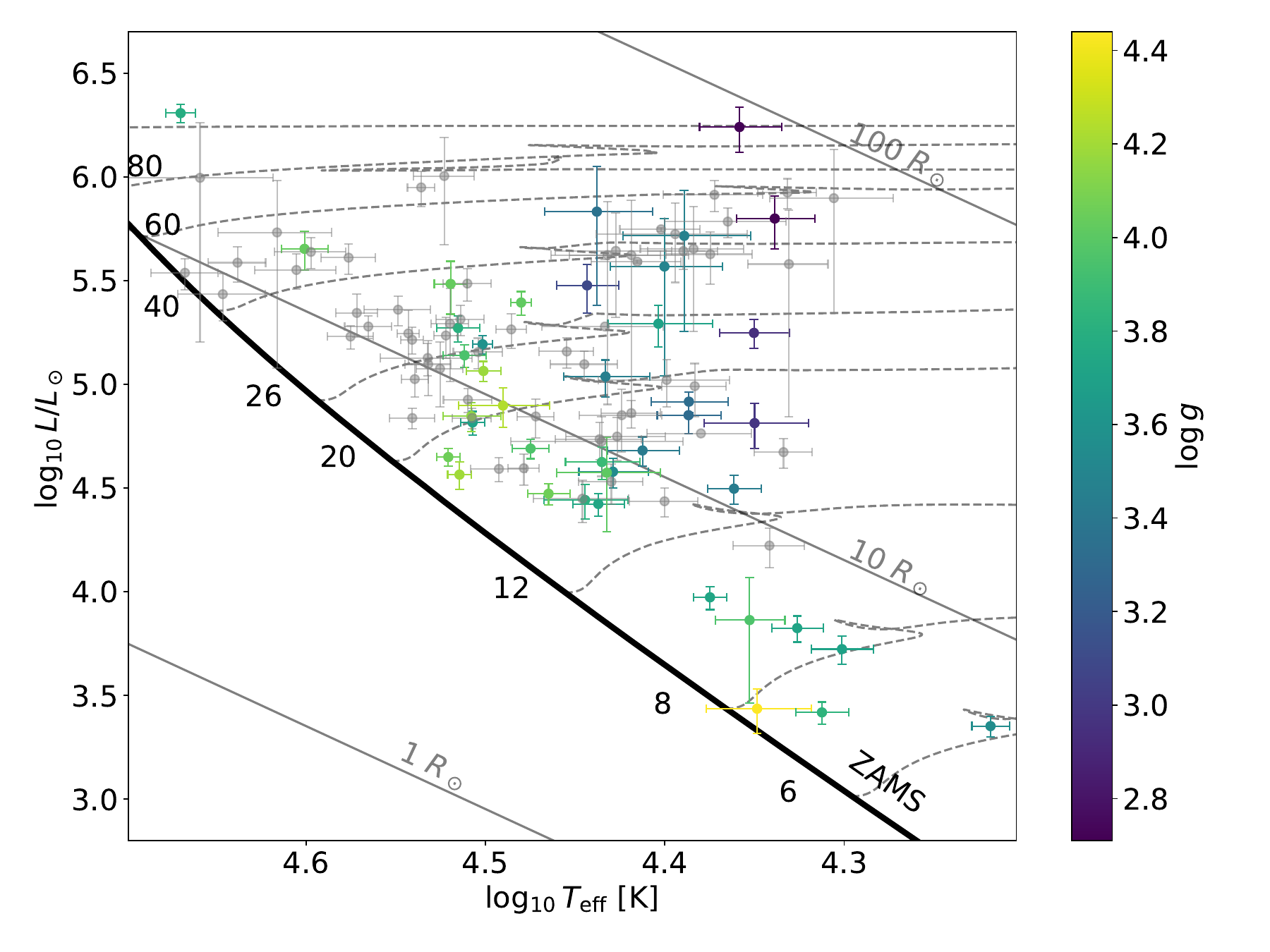}{0.72\textwidth}{(b)}}
        \caption{Spectroscopic (a) and conventional (b) HR diagrams for bowshock stars. Bowshock stars without apparent indications of binarity are color coded according to their measured \logg. Bowshock stars with indications of binarity are colored in gray. MIST evolutionary models are plotted in dashed lines, with ZAMS masses labeled.}
        \label{fig:HR}
\end{figure*}
 
Figure~\ref{fig:HR} presents a spectroscopic HR diagram\footnote{The spectroscopic HR diagram \citep{LangerandKudritzki2014} plots spectroscopic luminosity $\mathcal{L}$, defined as $\mathcal{L}=T_*^4/g\propto L_*/M_*$. The spectroscopic HR diagram is useful because it does not rely on a parallax or radius measurement and only requires stellar parameters obtained from spectra.} (a) and a conventional HR diagram (b) using luminosities derived from SED modeling. Solid diagonal tracks denote lines of constant gravity in the spectroscopic HR diagram (a) and constant radius in the conventional HR diagram (b). Bowshock stars that do not show apparent indications of binarity are colored by \logg~in both panels, and probable binary bowshock stars are represented in gray. MIST stellar tracks \citep{Choi2016}, computed with zero rotation and solar metalicity, are overplotted for a range of stellar masses in both panels in dashed gray lines. In the spectroscopic HR diagram (a), we expect binary bowshock stars, especially deblended SB2s, to not concord with stellar models as retrieved stellar parameters for these systems are questionable. In the conventional HR diagram (b), we expect the binary bowshock stellar luminosity to be the fitted luminosity of both components of the binary system.

In the spectroscopic HR diagram (a), we find bowshock stars show good agreement with evolutionary models for masses ranging from 6 -- 80~\msun. Many of them fall along the main sequence. Three high-\logg~stars stand out conspicuously in the spectroscopic HR diagram, below the zero-age main-sequence (ZAMS), just above 30~k\kelvin~and between the 12 and 20~\msun~models. These were flagged as being possible binaries.

In the conventional HR diagram (b), we again find bowshock stars agree with evolutionary models with masses ranging from 6--80~\msun. The low-mass, high-\logg~star in the bottom right of the spectroscopic HR diagram appears in a similar location as in the upper panel, and lies between the 6--8~\msun~tracks. The bowshock stars occupy the same locus on the upper panel as the lower panel, demonstrating the agreement between these two different luminosity indicators. In the bottom panel, high-\logg~stars cluster near the ZAMS, while low-\logg~stars scatter to the right. This is an expected result, as main-sequence stars have smaller radii at a given stellar mass than evolved stars. Probable binaries have larger mean uncertainties on luminsoity in both panels, as a result of the larger uncertainty on temperature in the spectral fitting. Larger temperature uncertainties propagate to the EXOFAST fitting, yielding large uncertainties in measure luminosity. 

\subsection{Are bowshock stars runaways?}

The formation mechanism of runaway stars is an active area of research. The two most prominent mechanisms are the Binary Supernova Scenario \citep[BSS;][]{Zwicky1957, Blaauw1961, Boersma1961} and the Dynamical Ejection Scenario \citep[DES;][]{Poveda1967, Leonard1991}. Some works propose a two-step mechanism in which both processes act on the same system when a dynamically ejected binary receives an additional kick as the primary undergoes a supernova \citep{Pflamm-AltenburgandKroupa2010, Dorigo2020}. In the BSS, the process of mass transfer from the primary to the secondary will spin up and deposit nuclear-processed material onto the secondary. With the reversal of the mass ratio, it is then expected the orbit will widen presupernova. The BSS will thus produce ``walkaway'' stars with high rotational velocities. \citet{Renzo2019} find, for example, a 90th percentile peculiar velocity of 20 \kms\ and that the mean runaway velocity of the secondary is largely independent of the parameters of the progenitor system. Interestingly, the BSS is thought to produce bound systems with the asymmetry in the SN explosions providing a kick to the entire system. In their simulations, \citet{Renzo2019} find 14\% of systems modeled resulted in a MS-compact object bound system. Such systems are thought to be origin of high-mass X-ray binaries \citep[HMXB;][]{Gott1971, Bolton1972, Webster1972, vandenHeuvel1972}. \citet{Phillips2024} proposed that mass transfer preceding the supernova of the primary is the source of Oe and Be stars, with the enhanced rotation creating a circumstellar decretion disk as the star surpasses its critical rotational velocity. The DES, on the other hand, is the process in which N-body interactions between massive stars and binary-binary interactions eject stars out of a natal cluster. The DES will leave its signature on the peculiar motions of the newly created runaway stars, with simulations finding binary-binary interactions creating runaways with speeds of the order of 100 \kms~\citep{Poveda1967}, with no enhancement on rotational velocity. Previous works have searched for indications of both processes to ascertain the prominence of each channel in the formation of runaways. \citet{Hoogerwerf2000} used proper motion data of nearby runaway stars and pulsars as indicators of the BSS and DES contributing to the production of runaways. They concluded that the high proper motion stars AE Aur and $\mu$~Columbae are likely to be products of the DES. Conversely, $\zeta$ Oph is a likely product of the BSS, traced back to a common locus in a young stellar group with oppositely directed, high-proper-motion pulsar J1932+1059. We expect to find two distinguishable runaway populations: BSS products with high projected rotational velocities and modest peculiar space veloicities; and DES products with low projected rotational velocities and high peculiar space velocities.

\begin{figure*}[htb!]
        \plotone{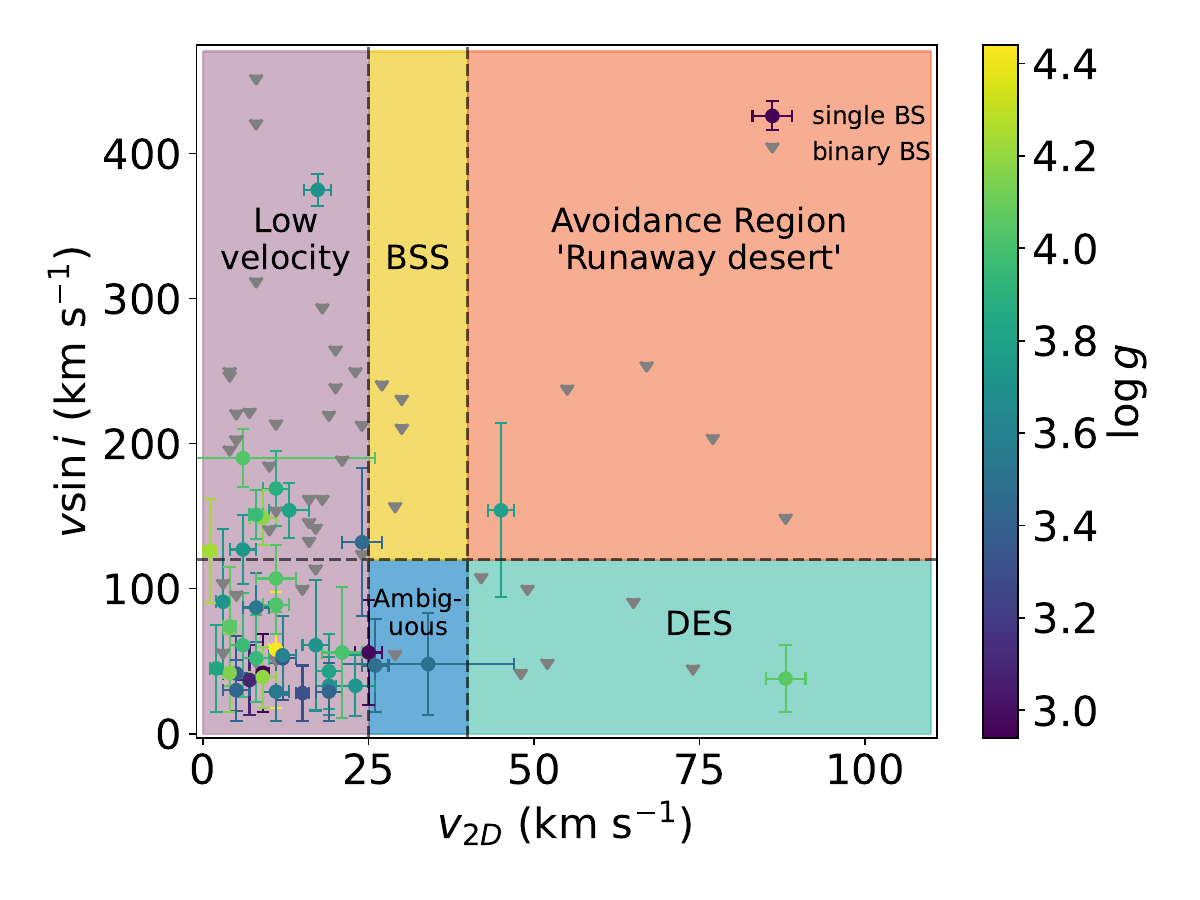}
        \caption{\vsini--peculiar space velocity distribution for bowshock stars. Single bowshock stars are colored by measured \logg, and binary bowshock stars are represent by grey triangles.}
        \label{fig:BSSDES}
\end{figure*}





Figure~\ref{fig:BSSDES} plots our projected rotational velocities against the 2-D peculiar space velocities, from \citet{KobulnickyandChick2022}, for bowshock stars. Single bowshock stars are colored by their measured \logg, and have error bars. Bowshock stars with indications of binarity are represented with grey triangles. For these systems, we expect the retrieved \vsini\ to represent an upper limit. We divided the plot into five regions: A low-velocity, non-runaway region with $v_{\text{2D}} <$25~\kms\ (\textit{purple infill}); the BSS region with projected rotational velocities $v\sin i>$120~\kms\ (well above the resolution limits of WIRO and KOSMOS spectrographs) and runaway velocity 25$<v_{\text{2D}}<$40~\kms\ (\textit{yellow}); the ambiguous runaway region with modest $v\sin i <$120~\kms\ and 25$<v_{\text{2D}}<$40~\kms\ (\textit{blue}); the DES region with $v\sin i<$120~\kms\ and $v_{\text{2D}}>$40~\kms\ (\textit{cyan}); and an avoidance region \citep[as noticed in the works][]{Sana2022, Guo2024} with $v\sin i>$120~\kms\ and $v_{\text{2D}}>$40~\kms\ (\textit{peach}). We find 19 of the 87 OB stars with kinematic data (22\%) are runaways under this criterion, consistent with the measured runaway fraction for OB stars measured in other studies \citep{Blaauw1961, Gies1986, Stone1979, Lamb2016}.

In the yellow BSS region there are four bowshock stars (BS$324_\star$, BS$362_\star$, BS$385_\star$, and BS$667_\star$), the first three of which present as double-lined spectroscopic binaries. The fact that we find an abundance of double-lined spectroscopic binaries in the BSS region is not surprising, considering blended spectra of two stars will be interpreted erroneously as rapid rotators. BS$667_\star$ (V1202 Sco) is a well-known eclipsing binary star \citep{Malkov2006, Avvakumova2013, Mellon2019}, and very likely is not rotating at the indicated \vsini of 150 \kms. In all likelihood, this star would be seen as a double-lined spectroscopic binary at higher spectral resolution. All four of the objects in this quadrant show indications of binarity. Therefore, the indicated \vsini\ in this region can be regarded as an upper limit and none of these objects are likely to be rapid rotators.

In the blue ambiguous region there are three stars (BS$019_\star$, BS$200_\star$, and BS$361_\star$), with BS$361_\star$ being classified as a single-lined spectroscopic binary. These stars are interpreted as either BSS stars with diminished projected rotational velocities due to the orientation of the star's rotational axis, or DES stars ejected with a velocity vector only partially in the plane of the sky.

In the cyan DES region of the plot there are seven stars (BS$039_\star$, BS$303_\star$, BS$343_\star$, BS$360_\star$, BS$375_\star$, BS$377_\star$, and BS$386_\star$), with BS$039_\star$, BS$343_\star$, and BS$377_\star$ classified as double-lined spectroscopic binaries, and BS$303_\star$ and BS$360_\star$ categorized as single-lined spectroscopic binaries. BS$375_\star$ (AE Aur), an O8--9 III, is classified as a DES product in this work, consistent with the conclusions of  \citet{Hoogerwerf2000}. BS$386_\star$ (B1 I) was determined to be a strong candidate for the BSS in \citet{Kaper1997} on the basis of being a component of an HMXB system.  It may fall into the DES region rather than the BSS region either because the star is viewed at low inclination or because the previously large rotation velocity diminished during its evolution into a supergiant. Alternatively, this object could be a member of a composite class of systems in which a binary ejected by N-body encounter subsequently receives an additional kick when the primary undergoes a supernova.

Five bowshock stars (BS$064_\star$, BS$339_\star$, BS$344_\star$, BS$353_\star$, and BS$383_\star$) fall into the peach avoidance region, with BS$064_\star$, BS$339_\star$, BS$353_\star$, and BS$383_\star$ classified as double-lined spectroscopic binaries. Given that these are erroneously classified as rapid rotators, it is reasonable that these four objects are actually runaway binary systems with lower rotational velocities. As for BS$344_\star$ (O6 If+), more commonly known as BD+43 3654, we find a best-fit \vsini\ of $145\pm43$ \kms, straddling the boundary of the DES and avoidance region. Other works \citep{Gvaramadze2011c, Carretero-Castrillo2023} determine N-body encounters to be the origin of this star's excessive peculiar velocity. Given these results and the imprecision in our measured \vsini, we conclude BS$344_\star$ should, in all likelihood, reside in the DES region.


Of the 19 bowshock stars that we deem as runaways, 15 show indications of binarity, suggesting a high incidence of binary bowshock runaways ($\geq$79\%). While both the BSS and DES mechanisms are thought to produce runaway bound systems, the efficiency of producing these systems varies. In their simulations, \citet{Renzo2019} find that of the stars that do not merge, 14\% of BSS systems remain bound after supernova of the primary while \citet{Perets2012} find the DES to produce runaways with a multiplicity fraction between 20--45\%. Given the abnormally high incidence of binarity in our sample of bowshock runaways, we consider this an indication of the DES being the preferred channel in producing runaway bowshock stars, and possibly runaway OB stars in general. 

\section{Conclusions} \label{sec:conclusions}

We have collected \numstars~low resolution blue spectra of bowshock-powering stars and measured best-fitting temperatures, gravities, and projected rotational velocities for the subsample of 103 OB stars. We have also obtained \numcomparison\ spectra for a sample of well-studied comparison OB stars and have shown that our analysis reproduces literature measurements of those stellar parameters. Archival optical and near-IR photometry provide constraints on additional stellar parameters radius, mass, luminosity, and visual extinction, to fully characterize the bowshock stars.










Bowshock stars span a range of temperatures from $16.5$~k\kelvin -- $46.7$~k\kelvin, \logg~from $2.50$--$4.60$, and projected rotational velocities $<100$--$400$~\kms, which are typical ranges of galactic OB stars. Notably, none of the bowshock stars have temperatures less than 16 k$K$ (excluding 2G3527642+0032660, an F dwarf that was erroneously selected as a candidate central star). Bowshock stars are unexceptional in their placement on both the spectroscopic and conventional HR diagrams. We determine 22\% of bowshock stars with kinematic data (19 of 84) are runaways, based on their two-dimensional peculiar space velocity ($v_{\text{2D}}>$25~\kms), consistent with other OB populations. We find \numbinaries~(\fracbinaries) of the bowshock stars show evidence of binarity. The high incidence of binary stars in runaway bowshock systems (15 of 19, $\geq$79\%) indicates that the DES is the preferred mechanism for creating bowshock runaways, and possibly OB runaways in general. With the exception of $\zeta$ Oph, no bowshock stars rotate at or near break-up speeds.


These basic data provide secure fundamental stellar parameters for 103 OB bowshock stars, in many cases for the first time. These basic data will inform future use of this sample to measure mass-loss rates \mdot~in the same manner as \citet{Kobulnicky2018, Kobulnicky2019}. Given our findings on runaway bowshock stars, we recommend future works investigate the characteristics of dynamically ejected- and binary supernova-ejected-systems, namely binarity, rotation rates, and peculiar space velocities for these populations.


\begin{acknowledgments}

This work has been funded by NSF through grants AST-2108347, AST-1852289 and AST-2108349.

\facilities{ARC, WIRO}

\software{IRAF (National Optical Astronomy Observatories 1999), EXOFASTv2: Generalized publication-quality exoplanet modeling code}

\end{acknowledgments}


\bibliography{bibliography}{}
\bibliographystyle{aasjournal}



\begin{appendix}\label{sec:appendix}

\section{Normalized spectra}
\label{sec:appendix-spectra}

\begin{figure*}[htb!]
        \plotone{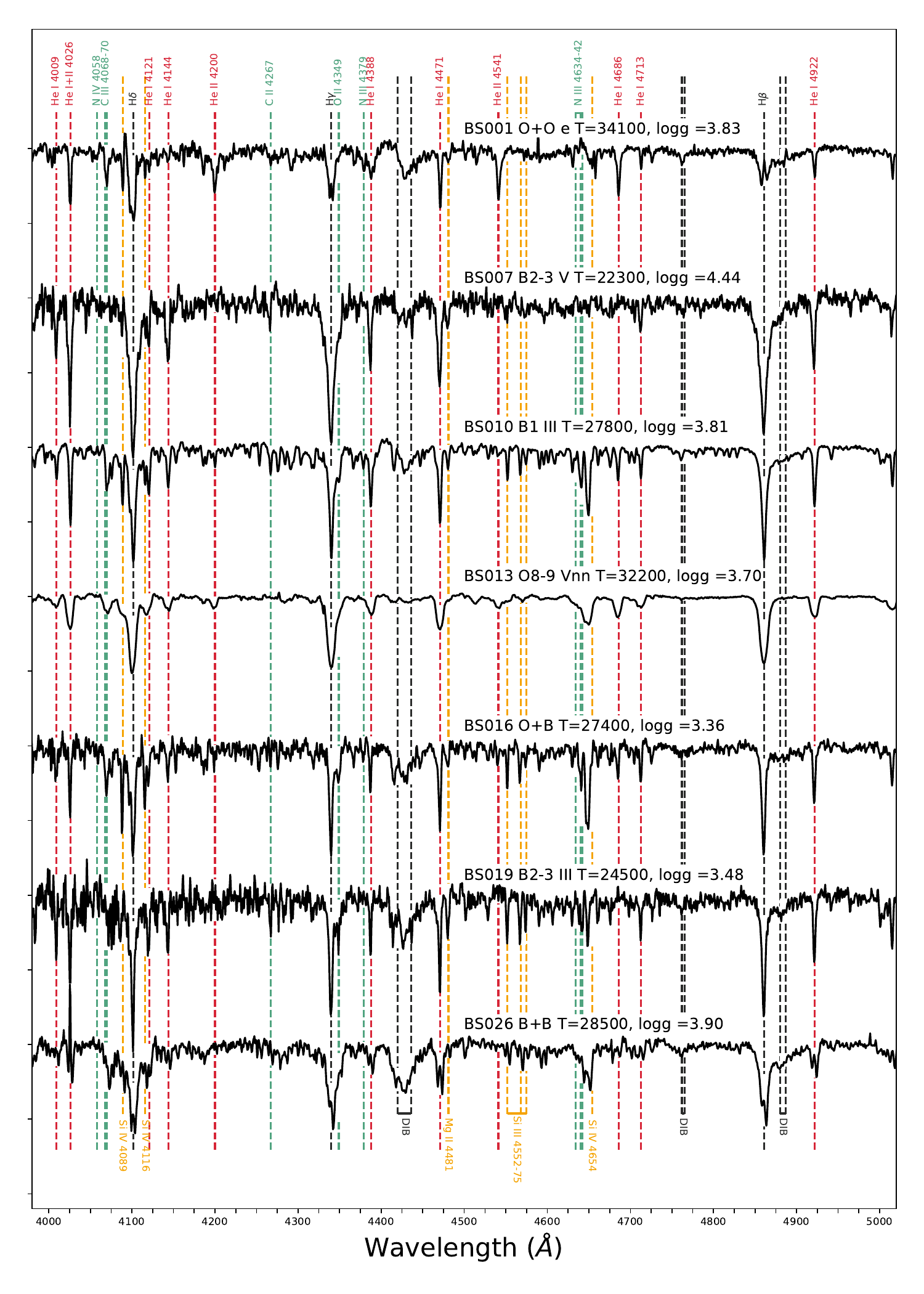}
        \digitalasset
        \caption{Bowshock stellar spectra}
        \label{fig:stack_spec0}
\end{figure*}

Figures in the figure set~\ref{fig:stack_spec0} present vertically stacked, normalized spectra for each bowshock star observed in this study on the wavelength range $\lambda\lambda$4000--5000~\angstrom, organized by bowshock identifier number up to BS709, then by galactic longitude for the \citet{Jayasinghe2019} objects. The complete figure set (15 images) is available in the online journal. We label characteristic absorption and emission features with their source atom and ionization state, and the approximate central wavelength in angstroms, in a similar style as the Galactic O-Star Spectroscopic Survey \citep[GOSSS,][]{Sota2011}. Balmer lines (\textit{top}) and diffuse interstellar bands (DIBs \textit{bottom}) are colored in black, neutral and singly ionized Helium (\textit{top}) are colored in red, Silicon (\textit{bottom}) in all ionizations in yellow/orange, and Carbon/Nitrogen/Oxygen (\textit{top}) in green. Trends in absorption strengths across temperature and surface gravities are often degenerate however for O- and B-stars, Helium generally indicates temperature and Silicon luminosity.

\figsetstart
    \figsetnum{16}
    \figsettitle{Vertically stacked normalized spectra of bowshock stars.}
    
    \figsetgrpstart
    \figsetgrpnum{16.1}
    \figsetgrptitle{Bowshock stellar spectra -- 1}
    \figsetplot{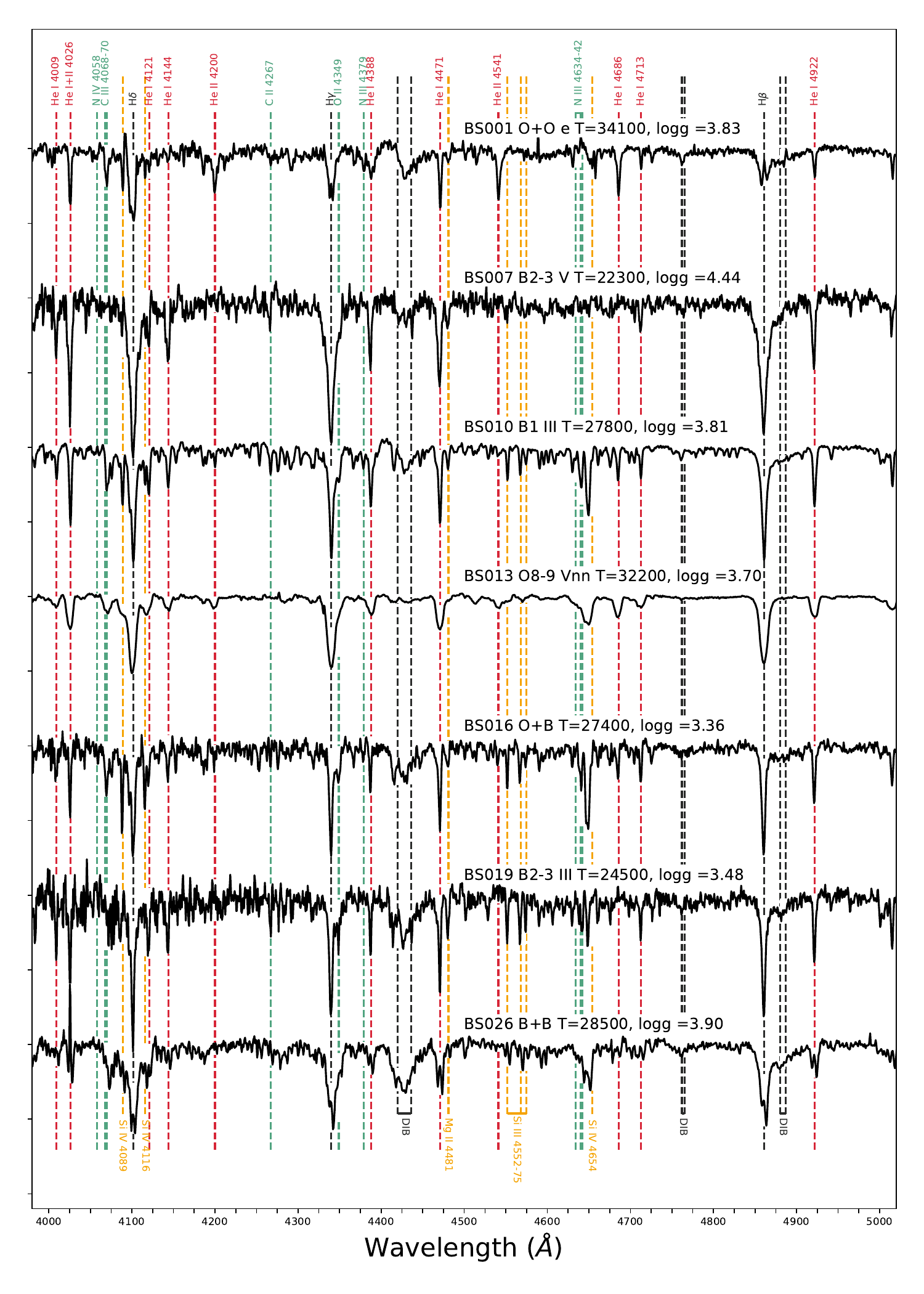}
    \figsetgrpnote{Bowshock stellar spectra}
    \figsetgrpend

\figsetend

\section{2G3527642+0032660 (HD 157642)}\label{app:HD157642}

The curious bowshock-powering candidate star 2G3527642+0032660 in \citet{Jayasinghe2019} (HD 157642, at a distance of 175~pc) may be a mistaken central star. HD 157642 is a bright star, with a spectral type F5--F8 \citep{Houk1982, CannonandPickering1993}. Our spectrum confirms this classification (based on the detection of \ion{Fe}{1} $4046$, $4383$ and \ion{Ca}{1} $4226$ features). It is likely therefore that another nearby star (or different physical phenomenon) is the source of the nebular emission surrounding HD 157642.

\begin{figure*}[htb!]
        \plotone{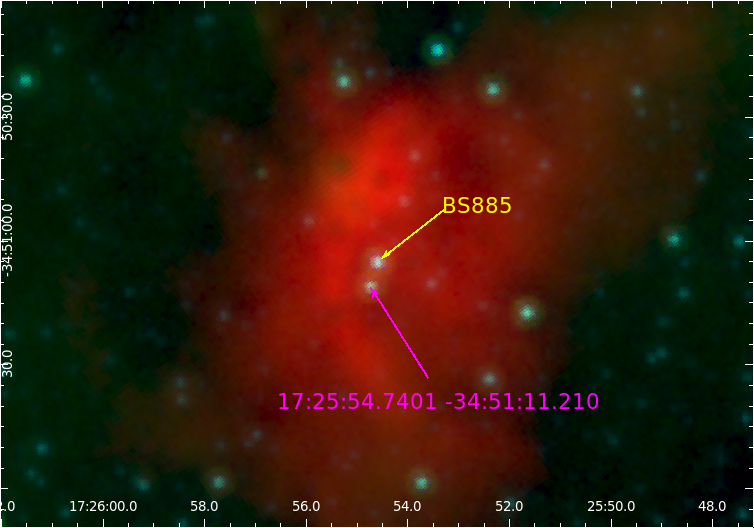}
        \caption{\textit{SST} image of HD 157642 (object 2G3527642+0032660 in the \citet{Jayasinghe2019} catalog) with red/green/blue representing $24$/$8.0$/$4.5$~\um~data.}
        \label{fig:IRHD157642}
\end{figure*}

Figure~\ref{fig:IRHD157642} presents the IR bowshock nebular surrounding HD 157642 using archival \textit{SST} data, with red/green/blue illustrating 24/8.0/4.5~\um~data, using IPAC \textit{SST} data explorer \citep{Spitzer}. The yellow arrow indicates HD 157642 (internal identification BS885) and the magenta arrow indicates the position of a nearby 4.5~\um~source. There is no Gaia source ID for this other source, suggesting that it is heavily extincted. It is possible therefore, assuming the nebula is indeed a stellar bowshock, that the heavily attenuated star could be the driving source of the bowshock nebulae and $2G3527642+0032660_\star$ is a foreground F5V star coincidentally along the axis of symmetry of the bowshock nebula.

\end{appendix}

\end{document}